\documentclass[a4paper,11pt]{article}
\pdfoutput=1 

\usepackage{jinstpub} 

\usepackage{longtable}
\usepackage{pdflscape}

\providecommand{\keywords}[1]{\textbf{\textit{JINST Keywords---}} #1}

\title{The KATRIN Superconducting Magnets: Overview and First Performance Results}

\affiliation[a]{Helmholtz-Institut f\"{u}r Strahlen- und Kernphysik, Rheinische Friedrich-Wilhelms Universit\"{a}t Bonn, Nussallee 14-16, 53115 Bonn, Germany}
\affiliation[b]{Institute of Experimental Particle Physics~(ETP), Karlsruhe Institute of Technology~(KIT), Wolfgang-Gaede-Str. 1, 76131 Karlsruhe, Germany}
\affiliation[c]{Institut f\"{u}r Physik, Johannes-Gutenberg-Universit\"{a}t Mainz, 55099 Mainz, Germany}
\affiliation[d]{Institute for Data Processing and Electronics~(IPE), Karlsruhe Institute of Technology~(KIT), Hermann-von-Helmholtz-Platz 1, 76344 Eggenstein-Leopoldshafen, Germany}
\affiliation[e]{Institute for Nuclear Research of Russian Academy of Sciences, 60th October Anniversary Prospect 7a, 117312 Moscow, Russia}
\affiliation[f]{Institute for Technical Physics~(ITeP), Karlsruhe Institute of Technology~(KIT), Hermann-von-Helmholtz-Platz 1, 76344 Eggenstein-Leopoldshafen, Germany}
\affiliation[g]{Max-Planck-Institut f\"{u}r Kernphysik, Saupfercheckweg 1, 69117 Heidelberg, Germany}
\affiliation[h]{Max-Planck-Institut f\"{u}r Physik, F\"{o}hringer Ring 6, 80805 M\"{u}nchen, Germany}
\affiliation[i]{Technische Universit\"{a}t M\"{u}nchen, James-Franck-Str. 1, 85748 Garching, Germany}
\affiliation[j]{Institute for Nuclear Physics~(IKP), Karlsruhe Institute of Technology~(KIT), Hermann-von-Helmholtz-Platz 1, 76344 Eggenstein-Leopoldshafen, Germany}
\affiliation[k]{Laboratory for Nuclear Science, Massachusetts Institute of Technology, 77 Massachusetts Ave, Cambridge, MA 02139, USA}
\affiliation[l]{Center for Experimental Nuclear Physics and Astrophysics, and Dept.~of Physics, University of Washington, Seattle, WA 98195, USA}
\affiliation[m]{Nuclear Physics Institute of the CAS, v. v. i., CZ-250 68 \v{R}e\v{z}, Czech Republic}
\affiliation[n]{Institut f\"{u}r Kernphysik, Westf\"{a}lische Wilhelms-Universit\"{a}t M\"{u}nster, Wilhelm-Klemm-Str. 9, 48149 M\"{u}nster, Germany}
\affiliation[o]{Department of Physics, Faculty of Mathematics und Natural Sciences, University of Wuppertal, Gauss-Str. 20, D-42119 Wuppertal, Germany}
\affiliation[p]{Department of Physics, Carnegie Mellon University, Pittsburgh, PA 15213, USA}
\affiliation[q]{Universidad Complutense de Madrid, Instituto Pluridisciplinar, Paseo Juan XXIII, n\textsuperscript{\b{o}} 1, 28040 - Madrid, Spain}
\affiliation[r]{Department of Physics and Astronomy, University of North Carolina, Chapel Hill, NC 27599, USA}
\affiliation[s]{Triangle Universities Nuclear Laboratory, Durham, NC 27708, USA}
\affiliation[t]{Commissariat \`{a} l'Energie Atomique et aux Energies Alternatives, Centre de Saclay, DRF/IRFU, 91191 Gif-sur-Yvette, France}
\affiliation[u]{University of Applied Sciences~(HFD)~Fulda, Leipziger Str.~123, 36037 Fulda, Germany}
\affiliation[v]{Department of Physics, Case Western Reserve University, Cleveland, OH 44106, USA}
\affiliation[w]{Institute for Nuclear and Particle Astrophysics and Nuclear Science Division, Lawrence Berkeley National Laboratory, Berkeley, CA 94720, USA}
\affiliation[x]{Institut f\"{u}r Physik, Humboldt-Universit\"{a}t zu Berlin, Newtonstr. 15, 12489 Berlin, Germany}
\affiliation[y]{Project, Process, and Quality Management~(PPQ), Karlsruhe Institute of Technology~(KIT), Hermann-von-Helmholtz-Platz 1, 76344 Eggenstein-Leopoldshafen, Germany    }

\author[a]{M.~Arenz,}
\author[b]{W.-J.~Baek,}
\author[c]{M.~Beck,}
\author[d]{A.~Beglarian,}
\author[j]{J.~Behrens,}  
\author[d]{T.~Bergmann,}
\author[e]{A.~Berlev,}
\author[f]{U.~Besserer,}
\author[g]{K.~Blaum,}
\author[h,i]{T.~Bode,}
% ...

\collaboration{\includegraphics[height=17mm]{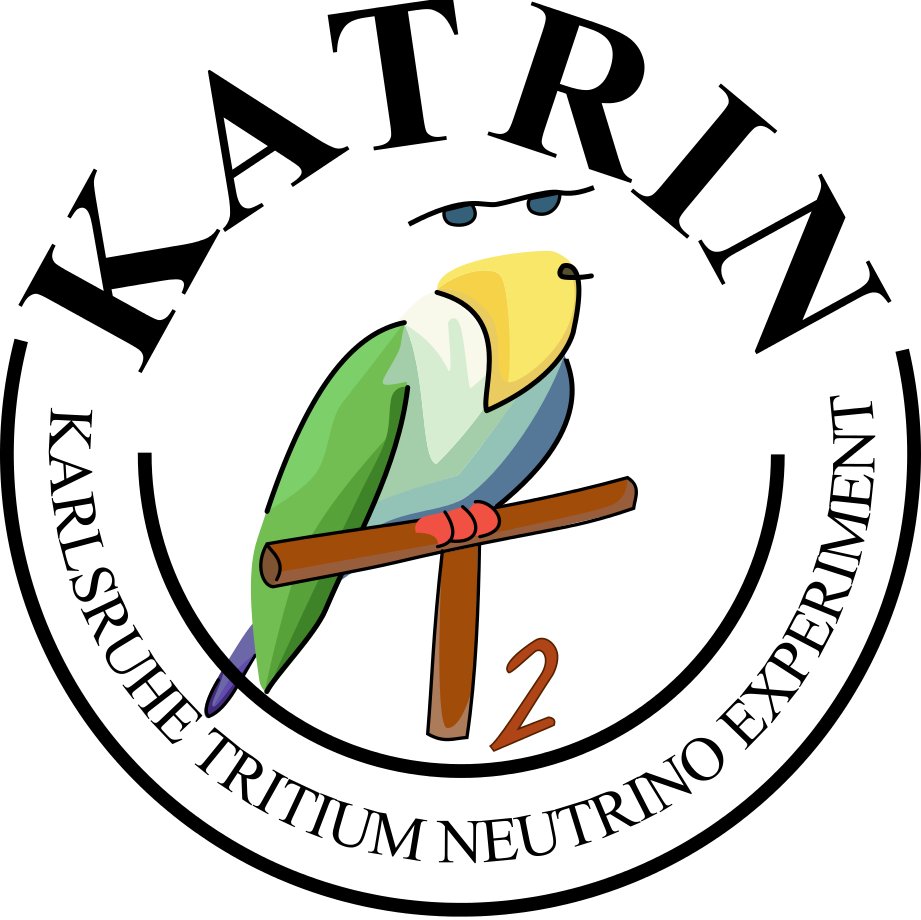}\\[6pt] The KATRIN collaboration}
\emailAdd{woosik.gil@kit.edu}

\abstract{The KATRIN experiment aims for the determination of the effective electron anti-neutrino mass from the tritium beta-decay with an unprecedented sub-eV sensitivity. The strong magnetic fields, designed for up to 6~T, adiabatically guide $\beta$-electrons from the source to the detector within a magnetic flux of 191~Tcm$^2$. A chain of ten single solenoid magnets and two larger superconducting magnet systems have been designed, constructed, and installed in the 70-m-long KATRIN beam line. The beam diameter for the magnetic flux varies from 0.064~m to 9~m, depending on the magnetic flux density along the beam line. Two transport and tritium pumping sections are assembled with chicane beam tubes to avoid direct ``line-of-sight'' molecular beaming effect of gaseous tritium molecules into the next beam sections. The sophisticated beam alignment has been successfully cross-checked by electron sources. In addition, magnet safety systems were developed to protect the complex magnet systems against coil quenches or other system failures. The main functionality of the magnet safety systems has been successfully tested with the two large magnet systems. The complete chain of the magnets was operated for several weeks at 70$\%$ of the design fields for the first test measurements with radioactive krypton gas. The stability of the magnetic fields of the source magnets has been shown to be better than 0.01$\%$ per month at 70$\%$ of the design fields. This paper gives an overview of the KATRIN superconducting magnets and reports on the first performance results of the magnets.}


\begin{document}
\maketitle
\flushbottom


\section{Introduction}
\label{sec:intro}
The determination of the absolute neutrino mass is of fundamental interest for particle physics and cosmology~\cite{KATRIN2005}. The Karlsruhe Tritium Neutrino (KATRIN) experiment aims for the determination of the effective neutrino mass (${m_{\bar{\nu_{e}}}}$) with a sensitivity of 0.2~eV/c$^2$ at 90$\%$~C.L. The measurement focuses on a energy region of several~eV around the endpoint ($E$~$\approx$~18.6~keV) of the tritium $\beta$-spectrum. The fraction of $\beta$-decays at the last eV before the end point is about 2$~\times~10^{-13}$~\cite{Drexlin2013}. This implies many technical challenges with respect to a high-luminosity tritium $\beta$-source, high energy resolution, and low background rates among others~\cite{KATRIN2005}. A factor of 10 improvement in mass sensitivity in comparison to former experiments~\cite{Kraus2005, Aseev2011} requires a factor of 100 increase in luminosity. 

The KATRIN experiment needs a chain of superconducting solenoid magnets (figure~\ref{fig1}) in order to guide the $\beta$-electrons from the source to the detector. Ten stand-alone single magnets and two large magnet systems were designed for the adiabatic stable transmission of $\beta$-electrons through the complete beam line. The installation of the complete chain of the magnets, with all beam tube sections, was finished in October, 2016. The first beam test from the source to the detector was successfully performed with a low-energy electron source on October 14, 2016~\cite{KATRIN2018a}. Further beam alignment tests were carried out with reduced magnetic fields during this first campaign. Recently, the complete chain of superconducting magnets were successfully operated at 70$\%$ of the maximum design fields for the first test measurements with radioactive krypton ($^{83m}$Kr) gas over a period of about three weeks~\cite{KATRIN2018a, KATRIN2018b}.

This paper gives an overview of the design of the KATRIN magnets with focus on the superconducting magnets and reports on their performance. The next section briefly describes the set-up of the KATRIN experiment which employs the MAC-E filter technique. Section~\ref{sec:ka-mag} explains the KATRIN superconducting magnets with details about each magnet system. In section~\ref{sec:mss}, magnet safety is described with focus on the protection of the two large superconducting magnet systems. Section~\ref{sec:commissioning} presents the first performance results of the superconducting magnets, followed by lessons learned in section~\ref{sec:lessons}. Finally, conclusions are drawn in section~\ref{sec:conclusion}.
\section{The KATRIN experiment}
\label{sec:katrin}
\begin{figure}[t]
\centering 
\includegraphics[width=150mm]{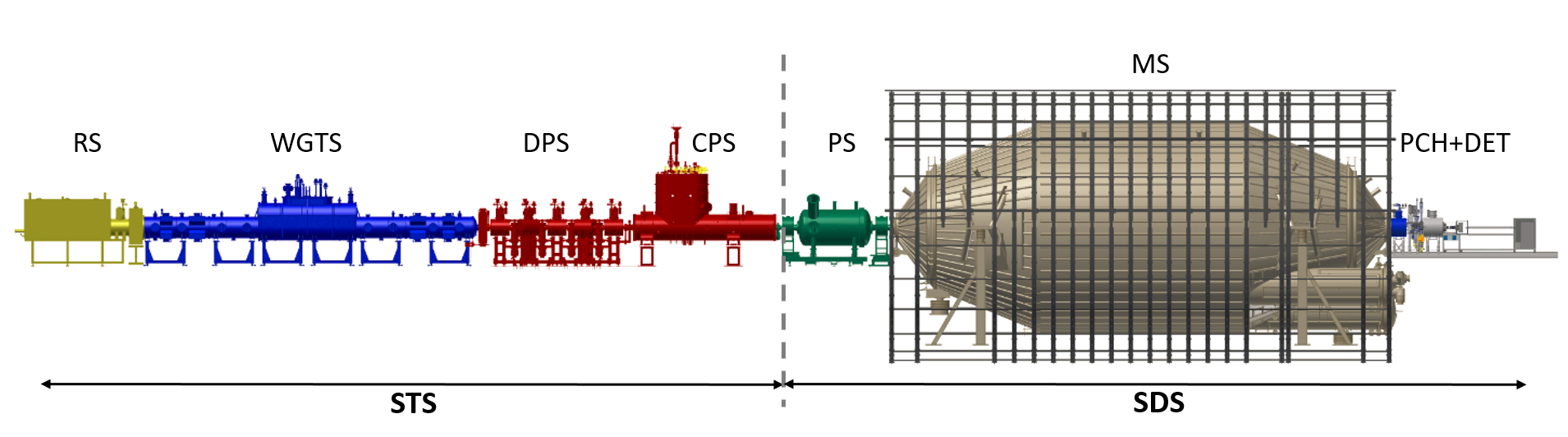} 
\caption{Overview of the 70-m-long KATRIN experimental set-up. RS: Rear Section, WGTS: Windowless Gaseous Tritium Source section, DPS: Differential Pumping Section, CPS: Cryogenic Pumping Section, PS: Pre-Spectrometer section, MS: Main Spectrometer section, PCH+DET: Pinch and Detector section, STS: Source and Transport Section, SDS: Spectrometer and Detector Section. The vertical dashed line indicates the wall between the STS and the SDS buildings. The components surrounding the MS indicate the air coil systems for magnetic field compensation at the analysing plane and fine-tuning of the magnetic flux density ($B_{min}$ in figure~\ref{fig1a}) (sections~\ref{sec:SDS} and \ref{sec:ms}), which is described in~\citep{Glueck2013, Erhard2017}.}
\label{fig1}
\end{figure}

The 70-m-long experimental set-up of the main KATRIN components is shown in figure~\ref{fig1} and has been installed at the Karlsruhe Institute of Technology (KIT) in Germany. The experiment is basically subdivided into two main sections. The first part is the Source and Transport Section (STS) containing the tritium-related components. The second part is the Spectrometer and Detector Section (SDS) with the non-tritium-related components. The two main sections are connected to each other by a DN~200 all-metal gate valve, which will be opened for data taking only.

At first, we briefly describe the measurement principle of the MAC-E filter technique, followed by the STS and the SDS. 
\subsection{MAC-E filter principle}
\label{sec:mac-e}
\begin{figure}[t]
\centering 
\includegraphics[width=150mm]{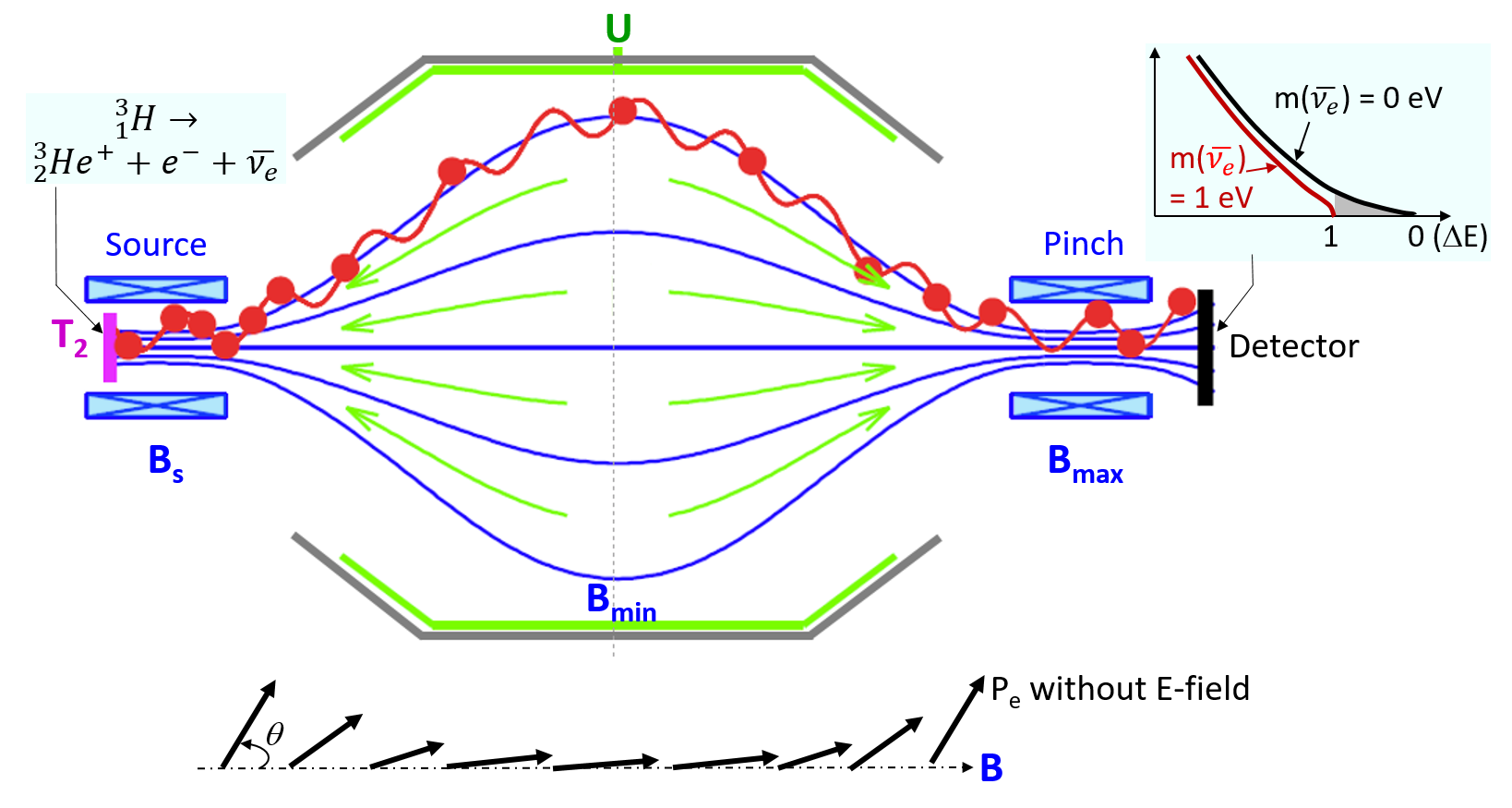}
\caption{Principle of the MAC-E filter. The $\beta$-electrons generated from the tritium beta decay at the source will be adiabatically guided by the magnetic fields ($B_{s}$, $B_{min}$, $B_{max}$) of the magnets to the detector. The electrostatic potential on the electrodes of the spectrometer $U$ generates the electrostatic retarding field (green arrow lines). The arrows at the bottom illustrate the adiabatic momentum transformation of the electrons with a polar angle $\theta$ in the inhomogeneous magnetic field. The polar angle $\theta$ at the source is the angle of the electron momentum relative to the magnetic field direction at the source. The plot on the right side illustrates the small difference of the $\beta$-spectra at the endpoint region $\Delta$E (= E$_s$ - E) for two cases: one with a neutrino mass ${m_{\bar{\nu_{e}}}}$~=~1~eV showing a `kink' at the endpoint and another one with zero neutrino mass without a kink.}
\label{fig1a}
\end{figure}

A key-component is the spectrometer using the so-called MAC-E filter (Magnetic Adiabatic Collimation with Electrostatic filter) technique for the precise analysis of the tritium $\beta$-spectrum. The MAC-E filter technique~\cite{Beamson1980} was well established in previous direct neutrino mass experiments~\cite{Lobashev1985, Picard1992, Kraus2005, Aseev2011}. The operating principle of a MAC-E filter is depicted in figure~\ref{fig1a}. The $\beta$-electrons with a kinetic energy $E_{s}$ from the source have to be adiabatically guided through the complete beam line to the spectrometer. They will move in a cyclotron motion along the magnetic field lines into the spectrometer. An adiabatic electron motion can be achieved according to eq.~\eqref{eq1a}\footnote{The expression eq.~\eqref{eq1a} is an approximation for the non-relativistic case.} by keeping the orbital magnetic moment ($\mu$) invariant during the transport in the magnetic field
\begin{equation}
\label{eq1a}
	\mu=\frac{E_{\bot}}{B} = const.,
\end{equation}
where $E_{\bot}$ is the transverse kinetic energy and $B$ the magnetic flux density. The adiabatic assumption of the electron motion is applicable if the magnetic field gradient is small during one cyclotron rotation: {$\nabla B/B~\ll$~1}.\footnote{The drift velocity of the centre of the electron cyclotron motion proportional to $\mathbf{E} \times \mathbf{B}$ and $\mathbf{B} \times \mathbf{\nabla_{\perp}}B$ is small compared to its orbital velocity~\cite{Jackson1975}} In the MAC-E filter, the initial transverse kinetic energy of the $\beta$-electrons at the source $E_{\bot,s}$ can be almost completely transformed into the longitudinal kinetic energy $E_{\|,a}$ at the analysing plane by continuously reducing the magnetic field strength, as shown in figure~\ref{fig1a}~(bottom).

The analysis of the $\beta$-spectrum can be accomplished by fine-tuning the electrostatic potential barrier $U$ at the analysing plane in the spectrometer, acting as an integrating high-energy pass filter. Only electrons with higher longitudinal kinetic energy than the electrostatic potential barrier can pass the filter and are re-accelerated and transported to the detector, where they are counted. Electrons with less kinetic energy are reflected. The relative sharpness of the energy resolution $\Delta E/E_s$ of the MAC-E filter is determined by the ratio of the minimum magnetic flux density $B_{min} = B_{A}$ at the analysing plane and the maximum magnetic flux density $B_{max}$ on the way to the detector with~\cite{KATRIN2005}
\begin{equation}
\label{eq1}
	\frac{\Delta E}{E_s} = \frac{B_{A}}{B_{max}}.
\end{equation}
This design value $\Delta E/E_s$ for KATRIN is 5~$\times~10^{-5}$ with $B_{A}$ = 0.3~mT and $B_{max}$ = 6~T. This allows a narrow filter width $\Delta E$~=~0.93~eV for the endpoint $E_s$~=~18.6~keV.

According to the magnetic mirror effect, the maximum polar angle $\theta_{max}$ (figure~\ref{fig1a}.) for the electron transmission to the detector is determined by the magnetic flux density at the source $B_s$ and the maximum magnetic flux density $B_{max}$~\cite{KATRIN2005}.
\begin{equation}
\label{eq2}
	\theta_{max} = \arcsin\sqrt{\frac{B_{s}}{B_{max}}}.
\end{equation}
For the designed field settings of KATRIN, $\theta_{max}$ is 50.8~$^\circ$ with a ratio of $B_{s}/B_{max}$ = 0.6. Electrons with a larger starting angle than $\theta_{max}$ will be reflected by the maximum pinch field $B_{max}$ before they reach the detector. Electrons with a smaller starting polar angle at the source than $\theta_{max}$ can be counted in the detector if their energy is large enough to pass the spectrometer. 
\subsection{The Source and Transport Section}
\label{sec:STS}
The STS is located in the Tritium Laboratory Karlsruhe (TLK) of KIT and comprises four main sections including tritium recycling loops: RS, WGTS, DPS, and CPS, shown in figure~\ref{fig1}. Gaseous tritium with a purity of 95$\%$ is supplied through the closed tritium circuits of TLK and monitored by the tritium diagnostic systems. The gaseous tritium is injected at a rate of about 40~g/day~\cite{KATRIN2005} in the middle of the 10-m-long central beam tube of the WGTS with a column density of 5$\cdot$10$^{17}$~T$_2$-molecules/cm$^2$~\cite{Kuckert2018}. This will generate about 10$^{11}$ $\beta$-electrons per second. The high flux of $\beta$-electrons must be adiabatically guided within a magnetic flux of 191~Tcm$^2$ at 100$\%$ of the design fields. The source properties with regard to its stability will be monitored by the RS on the rear side of the WGTS. More details of the RS are reported in~\cite{Drexlin2013, Babutzka2014}. The temperature of the 10-m-long central source beam tube, with a diameter of 90~mm, is stabilized in the WGTS by a two-phase neon cooling system at a temperature of about 30~K with a stability of 0.1$\%$ required for the KATRIN neutrino mass sensitivity. The temperature stability that has been achieved during test measurements was one order of magnitude better than the design value~\cite{Drexlin2013, Grohmann2008, Grohmann2009, Grohmann2013}. Gaseous tritium molecules in the WGTS which diffuse to the ends of the beam tube will be pumped out by 14 turbo-molecular pumps (TMP). 

Furthermore, the remaining tritium molecules will be pumped out by four additional TMPs in the DPS
~\citep{Gil2012, Kosmider2012} and by the cryo-sorption pump in the CPS~\citep{Gil2010, Jansen2015}. At the end of the STS, a tritium flow suppression factor of 10$^{14}$ has to be achieved, in order to limit the background rate in the spectrometer down to 10$^{-2}$~counts per second (cps). Some of the beam tube sections are inclined at an angle of 20~degrees in the 7-m-long DPS and 15~degrees in the 7-m-long CPS in order to improve the tritium pumping efficiency by blocking the direct ``line-of-sight'' molecular beaming effect to the next beam sections (figure~\ref{fig2}). This requires a precise assembly work with each beam tube section to achieve  unobstructed electron transport within the flux of 191~Tcm$^2$ through the complete beam line (Section~\ref{sec:beamcheck}). 

In parallel to the electron transport, most ions entering the DPS beam tubes will be eliminated by electrostatic dipole electrodes installed in the DPS beam tubes or blocked by ring electrodes~\cite{Klein2017}. The residual ions will be characterised by a FT-ICR (Fourier Transformation - Ion Cyclotron Resonance) trap in the last beam tube of the DPS~\cite{UbietoDiaz2009}. 
\subsection{The Spectrometer and Detector Section}
\label{sec:SDS}
The SDS is located in the spectrometer hall and includes two spectrometers (PS and MS) and the detector section for the $\beta$-electron measurement with the MAC-E filter. The 3.4-m-long pre-spectrometer (PS) with a diameter of 1.7~m will be operated as a pre-filter, reflecting all electrons with energies, for instance up to 300~eV below the end point, while allowing the endpoint part of the spectrum to be transmitted to the MS. The pre-filtering of lower energies can still be adjusted for searching, e.g., sterile neutrinos. The MS provides a high-energy resolution of 0.93~eV with a high voltage system, supplying up to 35~kV with ppm stabilities and absolute accuracy with precision high voltage dividers~\cite{Bauer2013a, Thuemmler2009}. The quality of the high voltage will be simultaneously monitored by the monitor spectrometer (MoS)~\cite{Erhard2014} which is installed in a separate hall and is not shown in figure~\ref{fig1}. There are two small superconducting magnets the MoS and were already used for the Mainz neutrino experiment~\cite{Picard1992} and are still in operation for the MoS. They will not be described in this paper. The MS has a diameter =~9.8~m, length =~23.2~m, inner surface area =~690~m$^2$, and a volume =~1240~m$^3$. The large diameter is needed to enclose the invariant magnetic flux of 191~Tcm$^2$ at a flux density of $B_A$~=~0.3~mT and to provide good adiabatic transmission conditions for the $\beta$-electrons. The fine-tuning of the magnetic field (B$_A$) at the analysing plane is performed by the individually controlled air-coil systems installed outside the MS vessel~(figure~\ref{fig1}), which compensate the earth magnetic fields and other distorting stray fields~\cite{Glueck2013}. Thus the magnetic field lines for the magnetic flux of 191~Tcm$^2$ in the MS can be contained inside the spectrometer vessel. The MS is also designed to maintain ultra-high vacuum conditions at 10$^{-9}$~Pa to minimize scattering of $\beta$-electrons~\cite{Wolf2016}. In addition, the Pinch magnet is located behind the exit of the MS and not before the MS to reduce background. The electrons with a larger starting angle in the source can be reflected in the spectrometers due to the magnetic mirror effect by setting the maximum magnetic field of the experiment at the exit of the MS according to eq.~(\ref{eq2}). More details about the challenges with the background reduction are addressed in a review article~\cite{Drexlin2013}. Once the $\beta$-electrons pass the centre of the MS, they will be accelerated towards the detector by the electrical fields in the MS and a 10~kV post-acceleration electrode inside the detector section to be counted by the detector. The focal plane detector consists of a monolithic silicon P-I-N diode wafer with 148 pixels covering a sensitive diameter of 90~mm. The detector wafer is located inside the warm bore of the detector magnet at 3.3~T with the design field configuration~\cite{Amsbaugh2015}. The sensitive pixels of the detector plane cover about 10$\%$ more flux than the magnetic flux of 191~Tcm$^2$ at the maximum design fields, allowing for a safety margin of the beam alignment.
\section{The KATRIN magnet systems}
\label{sec:ka-mag}
\subsection{Overview}
\label{sec:overview}
\subsubsection{Key design properties}
\begin{landscape}
\begin{figure}[t]
\centering 
\includegraphics[width=230mm]{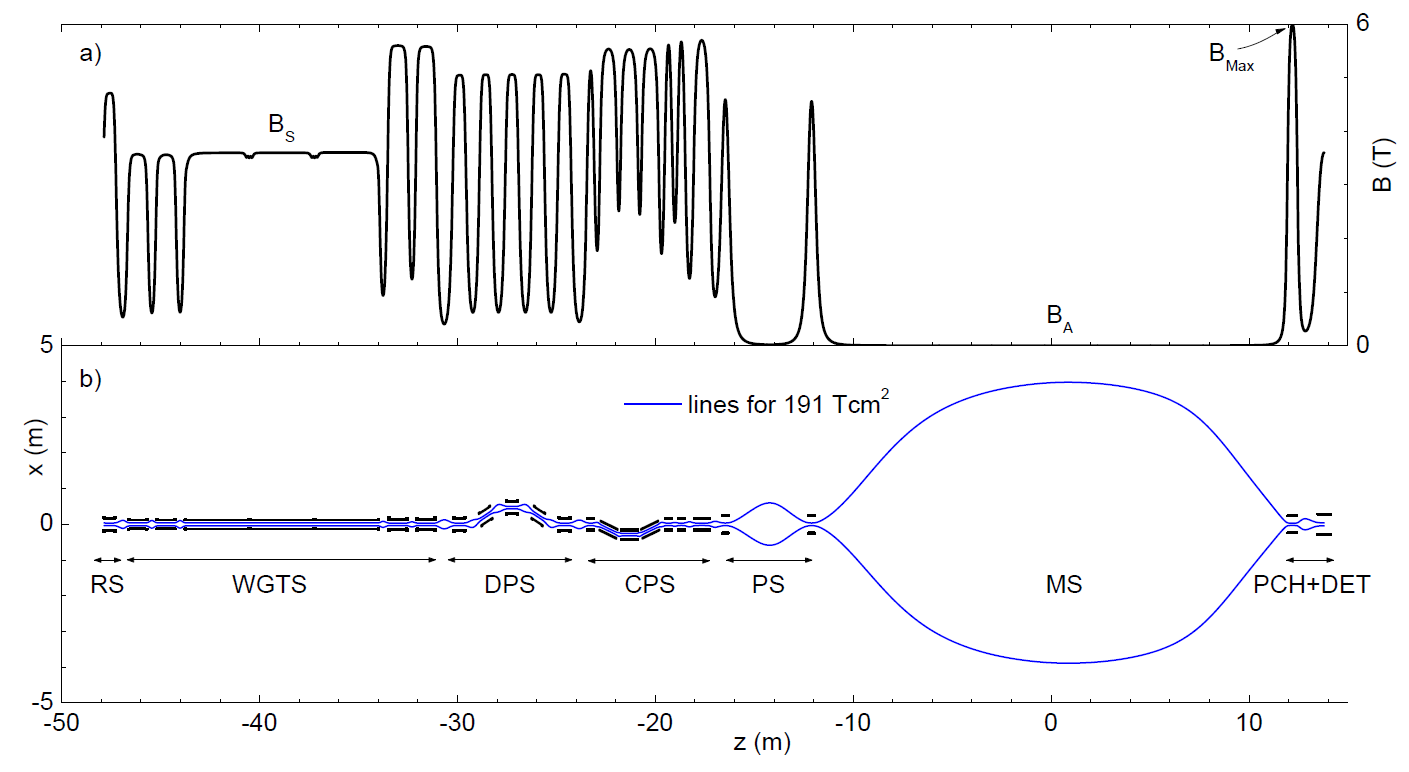}
\caption{a) Central magnetic flux density along the beam line and b) a geometrical overview of the complete chain of the KATRIN superconducting solenoids. $B_s$ is the magnetic flux density at the source, $B_A$ at the analysing plane of the MS, and $B_{max}$ at the Pinch magnet. The magnet modules in the DPS and the CPS are tilted by 20 degrees and 15 degrees relative to the z-axis, respectively. b) Two boundary lines calculated for a flux of 191~Tcm$^2$ at the 100$\%$ design fields are shown as an example of the guiding fields. Neither the dipole coil pairs of the WGTS nor the air coils of the MS are shown.}
\label{fig2}
\end{figure}
\end{landscape}

The KATRIN magnet systems are designed to provide the following key properties for the KATRIN experiment as introduced in the previous section:  
\begin{description}
\item[Magnetic flux of 191~Tcm$^2$ at 100$\%$ of the design fields] The chain of the magnet systems has to provide an invariant magnetic flux $\Phi$ for the adiabatic transmission of $\beta$-electrons through the complete beam line. The beam diameter $d$ for the constant magnetic flux ($\Phi$~=~191~Tcm$^2$) varies from 0.064~m to 9~m, depending on the magnetic field strength along the beam line because of $d = 2\sqrt{\Phi/\pi B}$ (figure~\ref{fig2}). The beam tubes must be aligned relative to the magnetic flux tube and provide sufficient clearance to avoid any interference with their inner structures such as inserts for ion reduction and monitoring~\cite{KATRIN2005}. For example, a beam tube diameter of 0.09~m is designed for the required flux diameter of 0.0822~m at the source. Especially challenging was the alignment of the tilted modules of the DPS and the CPS. Iterative magnetic field calculations had to be carried out to check clearances of the magnetic flux tube relative to the beam tube structures (Section~\ref{sec:beamcheck}).
\item[Maximum magnetic flux density ($B_{d}$)] The maximum magnetic flux density at the source and at the Pinch magnet is designed to reach high energy resolution and to restrict the maximum acceptance angle according to eq.~(\ref{eq1}) and eq.~(\ref{eq2}). The superconducting wires of the coil windings were specified with a proper safety margin against quenching, considering the peak fields in the coil windings, which are several percent higher than $B_d$.
\item[Stability of the magnetic flux density ($\Delta B/B_{d}$)] The magnetic guiding fields have to be stable in long-term operation, in order to minimize the systematic uncertainties of the experiment. The magnetic field variation $\Delta B/B_{d}$ should be below 0.03~$\%/$month at the source and at the Pinch magnet in order to keep the acceptance angle stable according to eq.~(\ref{eq2}). Drifts in the other transport magnets were specified with a value of 0.1~$\%/$month (Section~\ref{sec:stability}). 
\item[Homogeneity of the magnetic flux density] Electrons can be trapped by the magnetic mirror effect in areas with  inhomogeneous magnetic fields, in particular in the source. They will lose energy by scattering processes with the residual gases. Therefore, the magnetic fields over the 10-m-long central beam tube have to be as homogeneous as possible (Section~\ref{sec:wgts}).
\end{description}

The magnetic fields of the magnets were calculated to check the key magnetic properties during the design and the system assemblies. The code `Magfield3' was first developed in {C} language~\cite{Glueck2011b} for the magnetic field calculations and is now part of the {KASSIOPEIA} code developed for the study of electric and magnetic fields and the tracking of charged particles from sub-eV to keV energies~\cite{Furse2017}. The code is able to precisely calculate electron transmission properties with energy loss effects, such as synchrotron radiation and different scattering processes. The results have been qualitatively cross-checked with independent tools. For example, the magnetic field calculations, electron tracking and energy loss by the synchrotron radiation have been compared to the results of the software PartOpt~\cite{Osipowicz2014}.
\begin{landscape}
\begin{table}[t] 
\caption{Main design parameters of the KATRIN superconducting magnets. WGTS magnets are separated in three current circuits with WGTS-R, WGTS-C, and WGTS-F (figure~\ref{figW0a}). PM: persistent current mode, DM: driven mode, PSHTR: persistent switch heater; ''yes" means that a PSHTR is installed, ''no" for not-installed. $B_{max}$: maximum magnetic flux density at the centre of the magnet technically designed for test, $B_d$: designed magnetic flux density at the centre of the magnet for the experiment, $\Delta B/B_{d}$~($\%$/m.): stability of magnetic flux density in $\%$/month, $I_d$: design current for $B_d$, L: total inductance of the magnet group, $E$: stored magnetic energy of the group, $J$: current density on main coil, $t_{r}$: duration of magnet ramping to $B_{d}$, $N_{modules}$: number of modules, $N_{coils}$: number of coils, $d_w$: the bare diameter of superconducting wire (Cu/NbTi), $l_{coil}$: the length of the longest main coil, $d_i$: the smallest coil inner diameter, $d_b$: the smallest module bobbin diameter, $d_{BT}$: the smallest beam tube inner diameter.}
\label{tab1}
\centering
\vspace*{1ex}
\begin{tabular}{|l|*{14}{c|}}
\hline
Magnets&RS&\multicolumn{3}{c|}{WGTS}&\multicolumn{5}{c|}{DPS}&CPS&\multicolumn{2}{c|}{PS}&\multicolumn{2}{c|}{Detector}\\
\cline{3-5}\cline{6-10}\cline{12-13}\cline{14-15}
~&~&WGTS-R&WGTS-C&WGTS-F&M1&M2&M3&M4&M5&~&PS1&PS2&PCH&DET\\
\hline
Mode &PM&DM&DM&DM&PM&PM&PM&PM&PM&DM&DM&DM&PM&PM\\
PSHTR &yes&no&no&no&yes&yes&yes&yes&yes&no&yes&yes&yes&yes\\
$B_{max}$~(T) &5.5&3.6&3.6&5.6&5.5&5.5&5&5.5&5.5&5.6&4.5&4.5&6.0&6.0\\
$B_d$~(T) &4.7&3.6&3.6&5.6&5&5&5&5&5&5.6&4.5&4.5&6.0&3.6\\
$\Delta B/B_{d}$~($\%$/m.)&0.1&0.03&0.03&0.03&0.1&0.1&0.1&0.1&0.1&0.1&0.1&0.1&0.03&0.1\\
\hline
$I_d$~(A) &66.65&309.85&308.84&208.84&71.02&71.16&70.85&70.89&
71.16&200&156.39&155.41&86.98&56.15\\
$L$~(H) &291&27&34&75&288&290&290&290&289&172&76.3&77.4&427&647\\
$E$~(MJ) &0.65&1.30&1.60&1.62&0.73&0.73&0.73&0.73&0.73&
3.44&0.93&0.93&1.62&1.03\\
$J$~(A/mm$^2$)&144&151&151&100&155&155&155&155&155&
117&160&160&196&185\\
$t_{r}$~(hrs)&2.1&1.9&1.9&1.9&2.2&2.2&2.2&2.2&2.2&3.2
&1.7&1.7&4.5&1.9\\
\hline
$N_{modules}$ &1&3&2&2&1&1&1&1&1&7&1&1&1&1\\
$N_{coils}$ &3&9&6&6&3&3&3&3&3&7&1&1&3&1\\
$d_w$~(mm) &0.64&1.42&1.42&1.42&0.64&0.64&0.64&0.64&0.64&1.27&1.1&1.1&0.64&0.64\\
$l_{coil}$~(mm) &630&3064&3064&781&630&630&630&630&630
&913&320&320&489&698\\
$d_i$~(mm) &320&224&224&224&320&320&320&320&320&220&456&456&442&540\\
$d_b$~(mm) &254&208&208&212&254&254&254&254&254&203&400&400&346&448\\
$d_{BT}$~(mm) &78&90&90&90&76&76&76&76&76&75&200&200&150&150\\
\hline
\end{tabular}
\end{table}
\end{landscape}
\subsubsection{Modes of magnet operation}
\begin{figure}[t]
\centering 
\includegraphics[width=130mm]{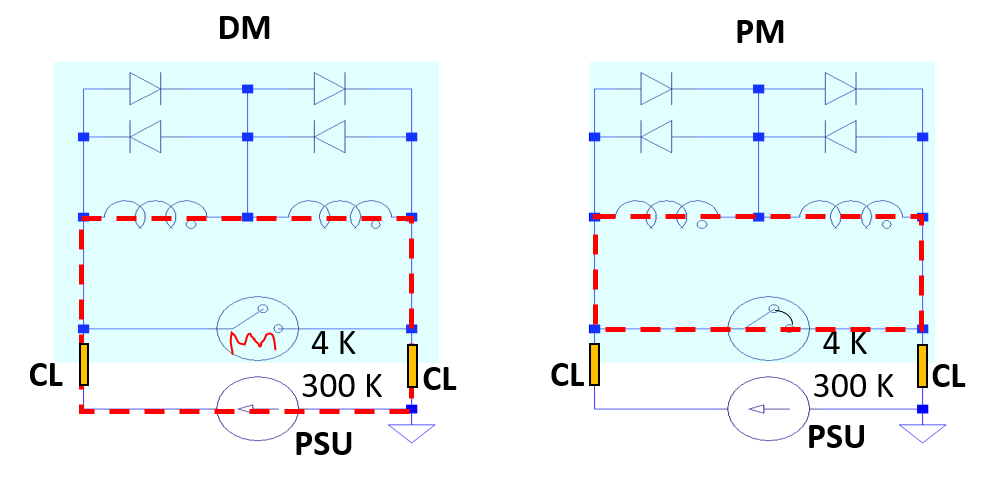}
\caption{Scheme of typical operation modes. Coils, diodes, and persistent switch are located at a cold state of 4~K. The red-dashed line indicates the coil current flow for each mode. In driven mode (DM) the coil current is driven by an external power supply unit (PSU) through the current leads (CL) (left). In persistent current mode (PM), the coil current is disconnected from the PSU and short-circuited by the persistent current switch (right). A free-wheeling cold diode pair is installed in parallel to each coil section for passive coil protection against quench.}
\label{figPM}
\end{figure}

Two typical operation modes can be considered for stable magnetic fields of the superconducting magnets: driven mode (DM) and persistent current mode (PM). Figure~\ref{figPM} shows a scheme of two typical operation modes. Magnets without persistent current switches can be charged by  external power supply units (PSU) and will be kept at the nominal current by the PSU in DM. The magnetic field stability of the magnets in DM is governed by the stability of the PSU. A magnet with a persistent current switch can be charged in driven mode by a PSU to the nominal current, after the persistent current switch has been opened by activating a persistent switch heater (PSHTR) (left). PM can then be set after reaching the nominal current by closing the persistent current switch by turning the persistent switch heater off (right). Small single magnets like NMR and MRI are typically operated in PM for a high magnetic field stability. Large complex magnets are preferentially operated in DM with stable power supplies, depending on the complexity of joints and the stability of the persistent current switch.   
\subsubsection{Short history of the KATRIN magnet design}
All superconducting magnets were conceptually specified by the KATRIN collaboration and were contracted to industrial partners for detailed design and manufacturing. There were several magnet design changes:
\begin{description}
\item[Number of magnet modules] At the beginning of the design, about 30 magnet modules in total were considered for eleven sections of the complete beam line~\cite{Gehring2004, KATRIN2005, Noe2006}. The two cryogen-free magnets of the PS were delivered in 2003 for early background investigations with the pre-spectrometer~\cite{Gehring2004}. However, the KATRIN beam sections were finally optimised with nine sections during the detailed design phase of the KATRIN experiment, requiring 24 superconducting magnet modules (figure~\ref{fig2}) from 3.6~T to 6~T. The magnets are cylindrical solenoids. Some magnet modules are wound with two compensation coils at their ends to allow an optimal magnetic flux and proper field homogeneity between neighbouring magnets. Therefore, the number of coils ($N_{coils}$) of the magnets is larger than the number of module ($N_{modules}$)~(table~\ref{tab1}). The initial Differential Pumping Section was ordered in 2003, the WGTS in 2004, and the CPS in 2008 from different industrial suppliers.
\item[Change of operation mode for the WGTS and the CPS magnet systems] PM was first specified for all superconducting magnets because of the high magnetic field stability for long-term operation. However, during manufacturing and test phases it turned out that the persistent current switch of the industrial partner for the source magnets was not stable for a current of 310~A. Furthermore, PM would demand such high quality of the final superconducting joint of the complex magnet assemblies as to pose a potential risk. Therefore, after discussions with external senior consultants, KATRIN decided to change the operation mode of the two large magnet systems - the WGTS and the CPS - from PM to DM with stable power supplies. The change of the operation mode from PM to DM additionally required the development of a proper magnet safety system (MSS) with external energy dumping units for protection (section~\ref{sec:mss}). In table~\ref{tab1} the operation modes of the KATRIN magnets are summarized, indicating the existence of a persistent switch heater (PSHTR).
\item[Cold bypass diodes for magnet protection] The former DPS~\cite{Gil2012} had to be replaced by five short single magnets (see the new DPS in figure~\ref{figDPS1}) because of unexpected damage to a cold bypass diode after a quench in 2011. The design of the cold bypass diodes for quench protection had to be improved for more reliability. The improved design also allows accessibility for an exchange of the bypass diodes, in case of a diode failure, without the need of a complex intervention in a system contaminated with tritium. The magnetic field direction within the WGTS and the CPS had to be fixed because of the bypass diodes which are designed for one polarity with regard to the global magnetic field direction of the KATRIN experiment, which in turn was defined \textit{against} the horizontal earth magnetic field in the spectrometer hall.
\end{description}
The chain of the KATRIN superconducting magnets was finally built with ten short single magnets (section~\ref{sec:rs},~\ref{sec:dps},~\ref{sec:ps}, and~\ref{sec:detector}) and two large complex magnet systems: the 16-m-long WGTS (section~\ref{sec:wgts}) and the 7-m-long CPS (section~\ref{sec:cps}). All superconducting coils were wound with low temperature superconductor Cu/NbTi round wires with twisted multi-filaments and with bare diameters from 0.64~mm to 1.42~mm. This type of superconducting wire is usually used for high magnetic field applications, such as NMR and MRI magnets. They were selected by the manufacturers, having a safety margin against quench between 10$\%$ and 30$\%$ along the load lines~\cite{Wilson1983}. The main data of the superconducting magnets are summarized in table~\ref{tab1}. In the following subsections we describe each magnet in detail. 
\subsection{Rear Section magnets}
\label{sec:rs}
\subsubsection{Description of magnets and operation mode}
The RS superconducting magnet is identical to the five new DPS magnets manufactured by Cryomagnetics, Inc. A 0.630-m-long main solenoid and two short correction coils on both ends of the main solenoid are combined to provide proper guiding fields and field homogeneities, as specified for the DPS single magnets in section~\ref{sec:dps}. The most important parameters of the magnet are summarized in table~\ref{tab1}. 

The magnet is passively protected against quenching by free-wheeling cold bypass diodes (figure~\ref{figPM}), as described in section~\ref{sec:mss}. An access port in the cryostat allows the replacement of the cold bypass diodes or the persistent current switch in case of damage. After the diode failure of the former DPS, this was an important design requirement for the new DPS and RS magnets. Otherwise, a later repair of the cold components would be very difficult once the beam line of the STS has been contaminated with tritium.
 
Ramping of the magnet to its design field of 5~T typically needs about 2.2~hours because of the large self-inductance of 291~H. However, this is not an issue for the experiment, because the superconducting magnet is operated in persistent current mode, providing a static magnetic field for a 60-day run cycle.

The magnet was successfully commissioned in 2015 at KIT with the five other single magnets of the DPS after a successful cold test at the manufacturer. In 2016, the RS magnet had a wire damaged; this resulted in a field drift larger than its design specification. It was successfully repaired and rechecked in September 2017~\cite{Gil2018a}. The present magnetic field stability of the magnet is reported in section~\ref{sec:stability}.

Behind the superconducting magnet of the RS, five small normal-conducting solenoids and four dipole pairs, with maximum fields of 50~mT and 0.3~mT respectively, are installed to allow steering of the electrons from an electron gun used for calibration purposes. Details of the normal-conducting coils are described with their dimensions in a thesis~\cite{Babutzka2014}.
\subsubsection{Instrumentation for magnetic field measurement}
\label{sec:rs_mag-sensor}
Two uni-axial Hall probes (Type~HHP-VP of AREPOC~s.r.o~\cite{Arepoc}) are installed on the outside of both end flanges of the magnet cryostat in order to monitor the magnetic field stability in persistent current mode during long-term operation. No magnetic field sensors were installed at the centre of the magnet because of the beam tube. However, the central magnetic field of the magnet set at its design field was measured with a NMR probe at the manufacturer site and re-checked with another NMR probe of METRO\textit{LAB} Instruments SA at KIT (section~\ref{sec:stability}). 
\subsubsection{Cooling system}
Magnet cooling is designed with the He-recondensing cryocooler system. The magnet cryostat is designed with an 80~K shield between the cryostat's outer shell and the first stage of the cold head and a 10~K shield between the first and second stages of the cold head, minimizing the heat load to below 1~W at 4.2~K. The superconducting coil is cooled in a 0.08~m$^3$ liquid helium bath by a two stage pulse-tube cryocooler Cryomech PT415. The cryocooler supplies a cooling power of 1.5~W at 4.2~K at the second stage and 40~W at 45~K at the first stage. The boiling helium can be easily recondensed by the second stage of the cold-head. The helium chamber is kept at a small over-pressure of about 4.8~kPa by regulating an electrical heater in the liquid helium bath. A small heating power between 0.3~W and 0.7~W indicates a recondensing cooling reserve. The boiling helium consumption of the recondensing magnet in persistent current mode is designed to be low enough for about nine months of continuous operation. The small helium consumption is typically associated with Joule heating on the normal conducting part of the current leads during magnet ramping in driven mode and with the leak tightness of the cryostat. A small number of sensors are installed for monitoring the magnet.
\subsection{Windowless Gaseous Tritium Source magnet system}
\label{sec:wgts}
\subsubsection{Description of magnets and operation mode}
\begin{figure}[t]
\centering 
\includegraphics[width=150mm]{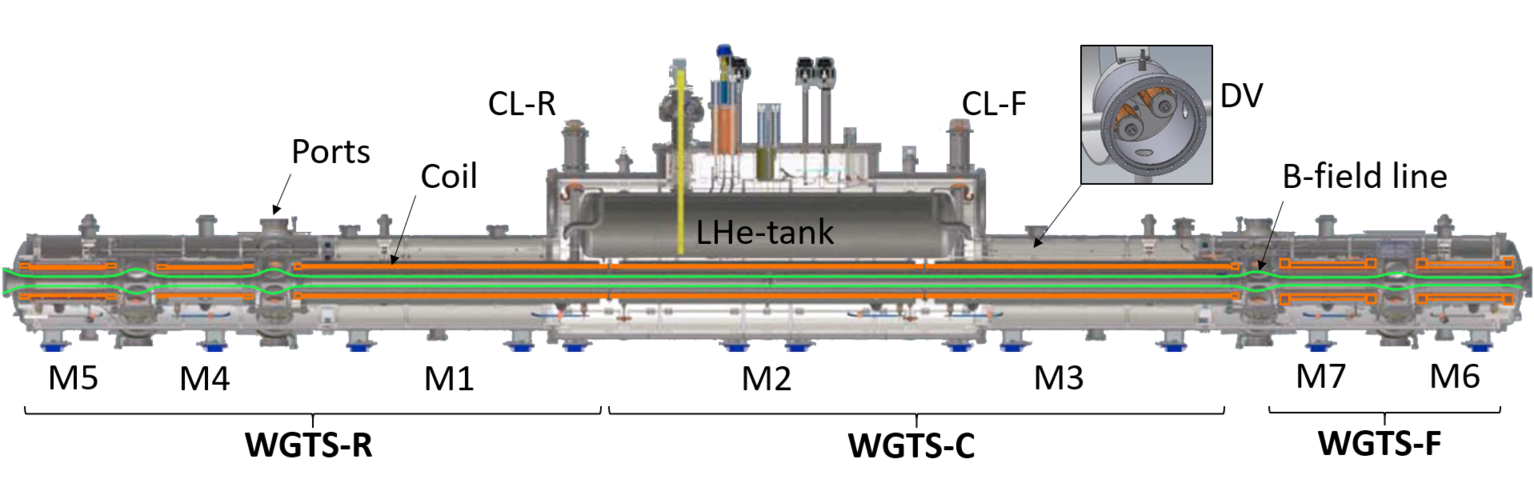}
\caption{Overview of the 16-m-long WGTS magnet system. Two dipole coil pairs are mounted on M5 and two other dipole coil pairs on M6, which are not shown. Two current-lead clusters are installed on the rear side (CL-R) and on the front side (CL-F) for all seven current circuits. An example of two diode vessels (DV) hosting diode stacks is shown in inset. Two magnetic field lines calculated for a magnetic flux of 191~Tcm$^2$ at the design fields are drawn. The magnetic fields decrease at the four pumping ports. The WGTS weighs about 27 tonnes.}
\label{figW0a}
\end{figure}

A schematic cross-section of the WGTS magnet system with seven solenoid modules is shown in figure~\ref{figW0a}. The 16-m-long magnet system is manufactured in one cryostat with a large liquid helium reservoir, long beam tubes, and pumping ports. Seven superconducting solenoid modules are installed in a straight line, surrounding five beam tube sections that are interconnected with four pumping ports. Three 3.3-m-long magnet modules (M1, M2, M3) are surrounding the 10-m-long central beam tube, connected with two 1-m- long magnet modules at both sides (M5 and M4 at the rear side, M7 and M6 at the front side). The dimensions of the WGTS magnets and their main data are listed in table~\ref{tab1}.

\begin{figure}[t]
\centering 
\includegraphics[width=150mm]{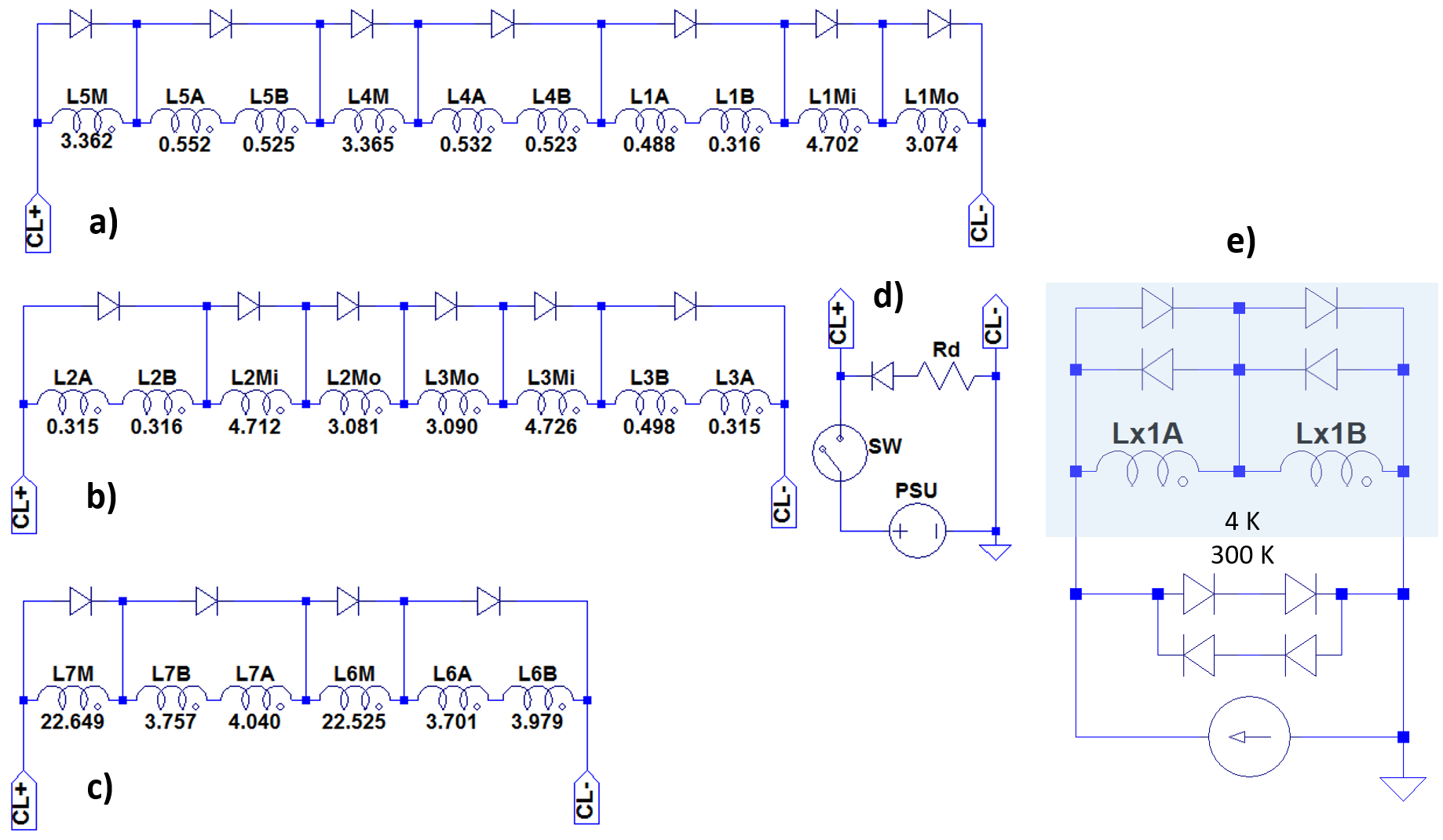}
\caption{Electrical scheme of the WGTS magnet circuits. The self-inductance of each coil segment is given  in Henries. The current circuits are shown for the magnet circuit of the coils of M5 (L5M, L5A, L5B), M4 (L4M, L4A, L4B), and M1 (L1A, L1B, L1Mi, L1Mo) of the WGTS-R (a), for the circuit of M2 (L2A, L2B, L2Mi, L2Mo) and M3 (L3A, L3B, L3Mi, L3Mo) of the WGTS-C (b), and for the circuit of M7 (L7M, L7A, L7B) and M6 (L6M, L6A, L6B) of the WGTS-F (c) with the external dumping unit (d). $R_d$ indicates the external resistance (table~\ref{tab3}). One typical electrical circuit of one pair of the dipole coils is also given with external free-wheeling diodes (e). Two vapour-cooled current-lead  clusters are installed on the rear side (CL-R) and on the front side (CL-F) for all seven different magnet circuits. A total of 33 cold bypass diodes are installed in parallel to each coil winding segment for magnet protection.}
\label{figWe1}
\end{figure}

Owing to the change of the operation mode from the PM mode to the DM mode, there are no persistent current switches installed. Thereby, the electrical current circuits of the seven main solenoid modules are optimised by reducing the stored magnetic energy of each circuit below 1.62~MJ by grouping the magnet modules in three groups; WGTS-R (M5, M4, M1), WGTS-C (M2, M3), and WGTS-F (M7, M6), as shown in the electrical scheme in figure~\ref{figWe1}. Thirty-three cold bypass diodes are installed in two separate diode vessels (DV) in form of diode stacks, as shown in figure~\ref{figW0a}(inset). However, protection of the magnets in driven mode is rather complicated because of non-negligible inductive couplings between the magnets. It requires a custom magnet safety system with external dumping units, as described in section~\ref{sec:mss}.

In addition, four dipole coil pairs are installed in order to deflect the main magnetic field lines in the x- and y- directions relative to the beam axes (z-direction) for the purposes of beam alignment and calibration. The maximum magnetic flux density of the dipole pairs are designed with 0.25~T at 110~A at the central beam axis, which is sufficient to deflect the guiding magnetic fields radially up to 42~mm. Two dipole coil pairs (DRx and DRy) are wound on the rear end module M5 of the WGTS-R, while two other pairs (DFx and DFy) are wound on the front end module M6 of the WGTS-F. 

The source magnetic field together with the field of the Pinch magnet is responsible for the maximum acceptable polar angle of the $\beta$-electrons according to eq.~(\ref{eq2}). The magnetic field has to be as homogeneous as possible over the 10-m-long central part of the beam tube, where gaseous tritium molecules will be injected and diffuse to both ends. The $\beta$-electrons generated in this central part of the source section will be mostly adiabatically guided by the magnetic field. 
\begin{figure}[t]
\centering 
\includegraphics[width=150mm]{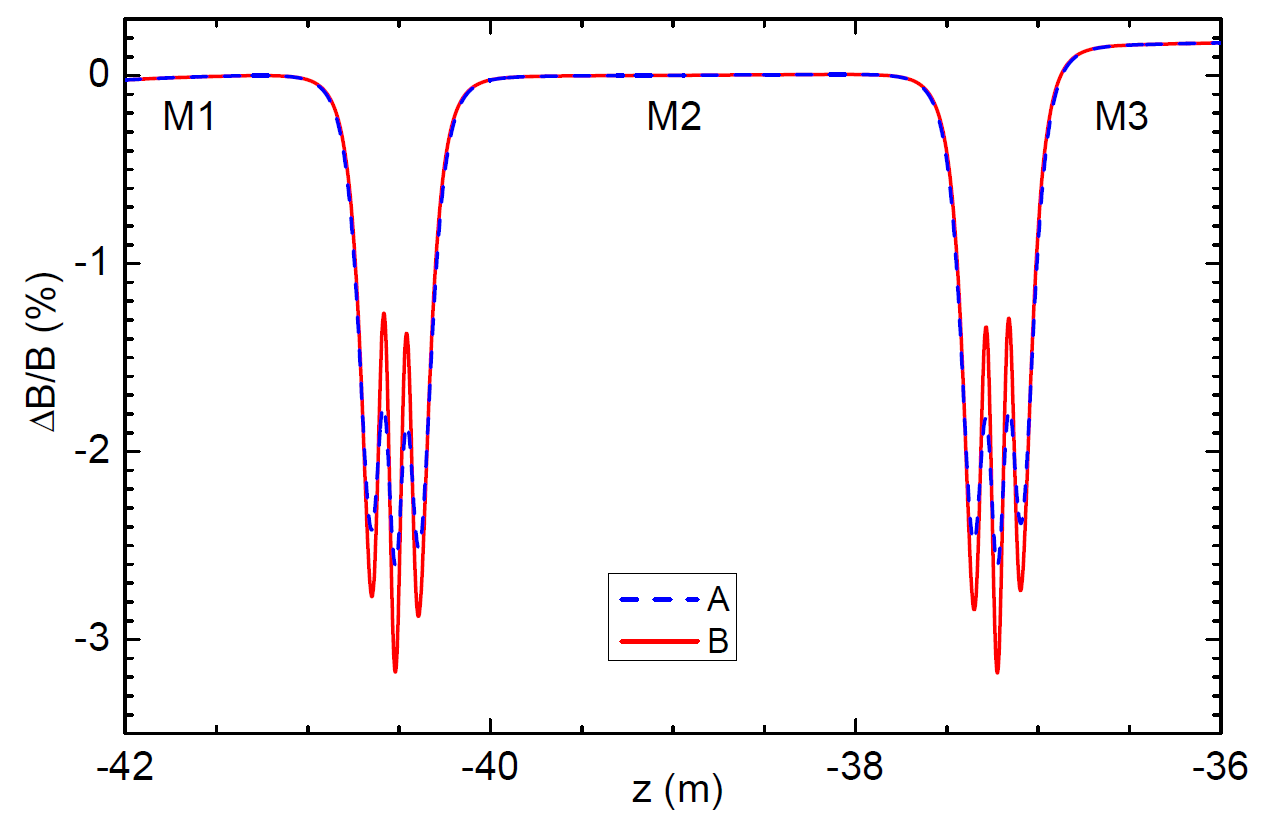}
\caption{Inhomogeneity of the magnetic fields in the WGTS. Two field calculations along the z-axis are shown for radius = 0 (A) and for the radius corresponding to 191~Tcm$^2$, as shown in figure~\ref{figW0a} (B). A small increase of about 0.2$\%$ in M3 is related to the slightly different coil winding numbers in the same current circuit of the WGTS-C.}
\label{figW2}
\end{figure}
\begin{table}[t]
\caption{Attracting magnetic forces in kN between the WGTS magnets and the neighbouring magnets at the design fields. ``x'' indicates the attracting magnet source. The three main groups of the WGTS are considered group-wise for simplification.}
\label{tab2}
\centering
\vspace*{1ex}
\begin{tabular}{|l|c|c|c|c|c|}
\hline
F$_{z}$~(kN)&RS&WGTS-R&WGTS-C&WGTS-F&DPS-M1\\
\hline
RS & x &15.0 &0.0&0.0&0.0\\			
WGTS-R& 15.0&x&200.9&0.0&0.0\\		
WGTS-C&0.0 &200.9&x&29.5&0.1\\
WGTS-F&0.0 &0.0 &29.5& x&14.8\\
DPS-M1&0.0&0.0&0.1&14.8&x\\
\hline
\end{tabular}
\end{table}
However, the $\beta$-electrons can be trapped in inhomogeneous magnetic field areas and lose energy by scattering processes with the residual gas. The construction of a 10-m-long solenoid system is technically challenging. Therefore, the central magnet was divided into three modules (M1, M2, and M3 in figure~\ref{figW0a}) each with a length of about 3.3~m after a detailed optimisation process~\cite{Gehring2006}. A small gap between these modules is unavoidable and defined by the mating flange thickness of the magnet chambers. The axial magnetic field inhomogeneity $\Delta B/B_{d}$ calculated with the coil data as wound is below 3.5$\%$ at the small gaps at both ends of module M2 (figure~\ref{figW2}). It is one order of magnitude smaller in the centre of the solenoid module.

On the other hand, because of the short separation distance between the long magnet modules, the maximum magnetic force between them is very high, 200.9~kN (table~\ref{tab2}), which had to be taken into account in the mechanical design. The energizing of the WGTS magnets has to be synchronized with the neighbouring magnets because of the strong inductive coupling between the three long magnet modules.

Most of the WGTS was designed by the former company ACCEL. Single modules were manufactured by the company. After the magnet modules were cold tested to their full fields at CEA and at Bruker-BASC, the WGTS was then finally assembled by RI~Research Instruments GmbH and KIT. It was delivered to KIT in September, 2015.
\subsubsection{Instrumentation for magnetic field measurement}
The magnetic field drift at the source has to be below 0.03~$\%/$month. Two uni-axial Hall probes (Type~HHP-VP of AREPOC~s.r.o~\cite{Arepoc}) are installed on the flange of one compensation coil of each module inside the module chamber for monitoring the magnetic fields. However, the stability of the magnetic fields can also be monitored directly with a current transducer DCCT manufactured by LEM$^\circledR$ in a higher precision than the Hall probes, because the magnets are operated in driven mode. A closed-loop fluxgate sensor IT 400-S Ultrastab~\cite{LEMit400s} manufactured by LEM$^\circledR$ is installed on each current circuit of the magnets outside of the cryostat. It has an accuracy of 0.0044$\%$. Two air-cooled PSUs of type NTS~2450-7~MOD manufactured by FuG~Elektronik~GmbH are used for the WGTS-R and the WGTS-C with a maximum current of 350~A. Another DCCT sensor ITN 600-S Ultrastab~\cite{LEMitn600s} manufactured by LEM$^\circledR$ is installed inside each PSU, which has a better accuracy of 0.00173$\%$. An air-cooled PSU of type NTS~2500-10~MOD from FuG~Elektronik~GmbH is adapted for the WGTS-F with a maximum current of 250~A. The current is measured with a 2 m$\Omega$ shunt. The current stability of the PSUs was specified to be better than 10~ppm per 8~hours. The results of the current stability tests with the magnets are reported in section~\ref{sec:stability}. 
\subsubsection{Cooling system}
The superconducting coils of the WGTS and of the CPS are cooled in liquid helium bath at 4.5~K and 0.13~MPa. A TCF 50 refrigerator of LINDE KRYOTECHNIK AG with a cooling power of 450~W supplies supercritical helium at 5~K with 0.5~MPa, which is distributed to the valve boxes of the WGTS and the CPS via 20-m- and 40-m-long cryogenic transfer lines, respectively~\cite{Grohmann2010}. Each cryostat of the WGTS and of the CPS is connected from the valve box through several-meters long flexible transfer lines for five He process lines (a 5~K supply line, a 300~K supply line for mixing helium gas with cold helium, two return lines for 5~K He and for He below 100~K to the cold box of the refrigerator, and a return line for He warmer than 100~K through the water bath heater back to the compressor). The supercritical helium is then liquefied into the He reservoir at 0.13~MPa by a Joule-Thomson expansion valve. The volume of the He reservoir is about 1.5~m$^3$ for the WGTS and 1.3~m$^3$ for the CPS. The total volume of the liquid helium inventory including all volume of the magnet chambers is about 2.8~m$^3$ for the WGTS and 1.6~m$^3$ for the CPS.

The cooling of the two-phase beam tube and other components has been described in~\cite{, Grohmann2008, Grohmann2009, Grohmann2010, Grohmann2013}.
\subsection{Differential Pumping Section magnets}
\label{sec:dps}
\subsubsection{Description of magnets and operation mode}
\begin{figure}[t]
\centering 
\includegraphics[width=150mm]{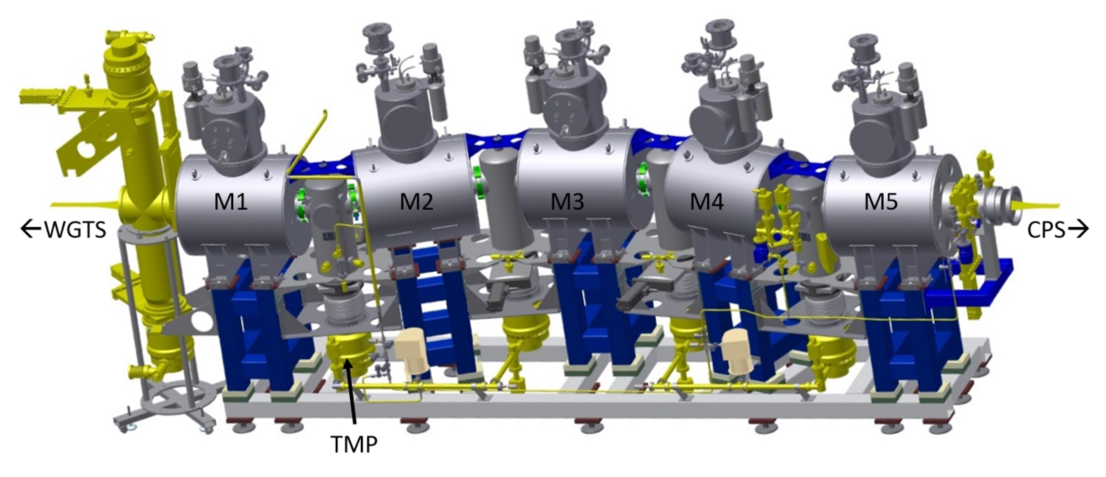}
\caption{Picture of the DPS with five single magnets. Turbo-molecular pumps (TMP) are installed on each pumping port.}
\label{figDPS1}
\end{figure}

As referred to section~\ref{sec:rs}, the five new DPS superconducting magnets were manufactured to the same design as the RS magnet by the same company. They are operated in the persistent current mode. A picture of the 3D-model of the five single magnets of the new DPS is shown in figure~\ref{figDPS1}. The DPS magnets are arranged with an angle of 20 degrees to each other with a large distance for pumping ports of a turbo-molecular pump (TMP) DN250 MAG2800W for tritium pumping efficiency. Each magnet was carefully aligned on its support structure using a laser survey and FARO arm measurements~\cite{FARO}. These measurements were then compared to the global magnetic field simulation in order to avoid any interference of the magnetic flux inside with the beam tubes and the pumping ports~\cite{Sack2015}. 
A magnetic field radial homogeneity of 100~ppm is designed around 40~mm from the magnet centre. But the homogeneity is reduced to about 1000~ppm in operation with neighbouring magnets.
 
The magnets were successfully cold-tested either alone or in triplet arrangement to check the maximum axial magnetic force of about 33~kN at 5.5~T to each other and the radial force of about 6~kN. All five modules were also successfully commissioned to 5.5~T at KIT.
\subsubsection{Instrumentation for magnetic field measurement}
Two uni-axial Hall probes (Type~HHP-VP of AREPOC~s.r.o~\cite{Arepoc}) are installed on the outside of both end flanges of each magnet cryostat for monitoring the magnetic field stability for long-term operation, as mentioned in section~\ref{sec:rs_mag-sensor}. 
\subsubsection{Cooling system}
The cooling system of each DPS magnet is identical with the one of the RS magnet, as already described in section~\ref{sec:rs}.
\subsection{Cryogenic Pumping Section magnet system}
\label{sec:cps}
\subsubsection{Description of magnets and operation mode}
\begin{figure}[t]
\centering 
\includegraphics[width=150mm]{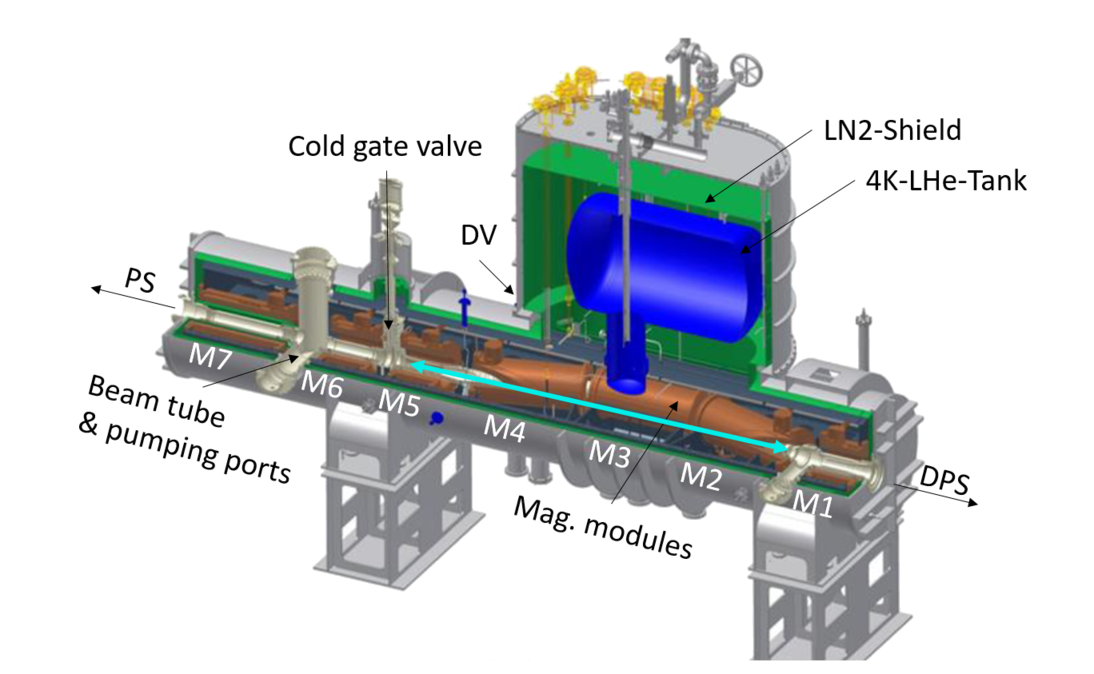}
\caption{Overview of the CPS magnet system with seven magnet modules, M1 to M7. The arrow from M2 to M5 indicate the cryo-sorption tritium pumping area on the beam tube sections at 3~K for standard operation. ''DV" indicates the location of the diode vessels (figure~\ref{figCd1}) . They are not visible in this model. The CPS weighs about 13 tonnes.}
\label{figC0a}
\end{figure}

\begin{figure}[t]
\centering 
\includegraphics[width=150mm]{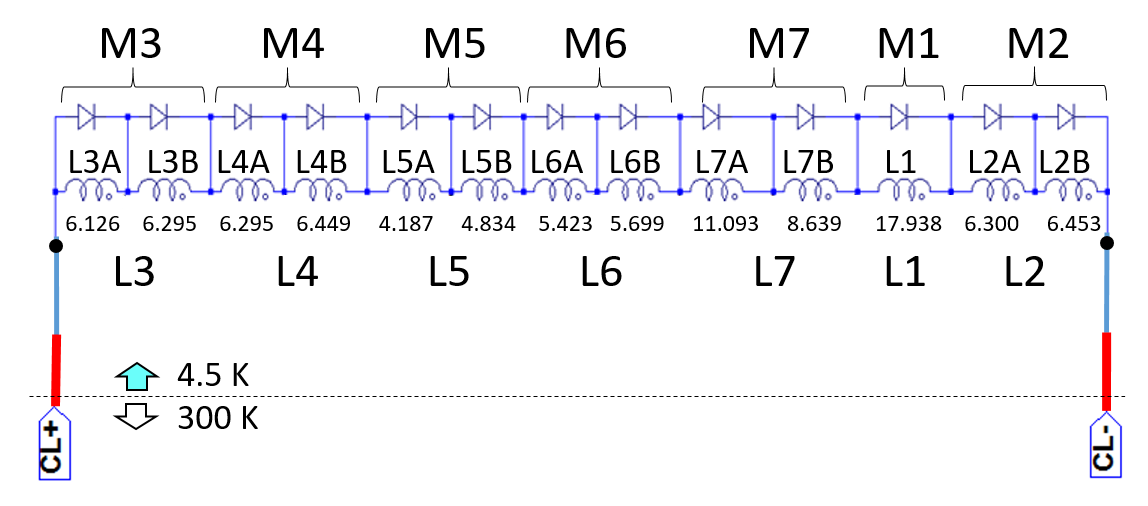}
\caption{Electrical scheme of the  CPS coils. A power supply unit and a dumping unit which are not shown here are connected to the current leads (CL), as shown in figure.~\ref{figWe1}d. ``Ln'' denotes \textit{n-th} module. ``A'' denotes the inner winding segment of ``Ln'' and ``B'' the outer winding segment of ``Ln''.  The self-inductance of each coil segment is given in Henry.}
\label{figCe1}
\end{figure}

An overview of the 7-m-long CPS magnet system with seven solenoid magnet modules from M1 to M7 is shown in figure~\ref{figC0a}. The CPS magnet system is housed in one cryostat because of the cooling of the beam tube for the cryo-sorption of tritium at 3~K~\cite{Gil2010}. Each solenoid coil (Ln with n=1 to 7~in figure~\ref{figCe1}) is housed in its helium chamber of the magnet modules (Mn from n=1 to 7). The seven magnet modules are assembled on the cold support structures inside of the cryostat vessel. The three modules from M2 to M4 are installed at a short separation distance each other because of the 15-degree chicane.

The complex CPS magnet system is designed to be operated in the driven mode, like the WGTS, avoiding a risk with persistent switch. The seven modules are driven in one electrical current circuit with a stable power supply (figure~\ref{figCe1}). 13 cold bypass diodes are installed in parallel to each coil winding segment for magnet protection, as shown in figure~\ref{figCe1}. Six solenoids from M2 to M7 are wound in two coil segments, divided into inner and outer windings, while M1 is wound without segmentation. 

\begin{figure}[t]
\centering 
\includegraphics[width=150mm]{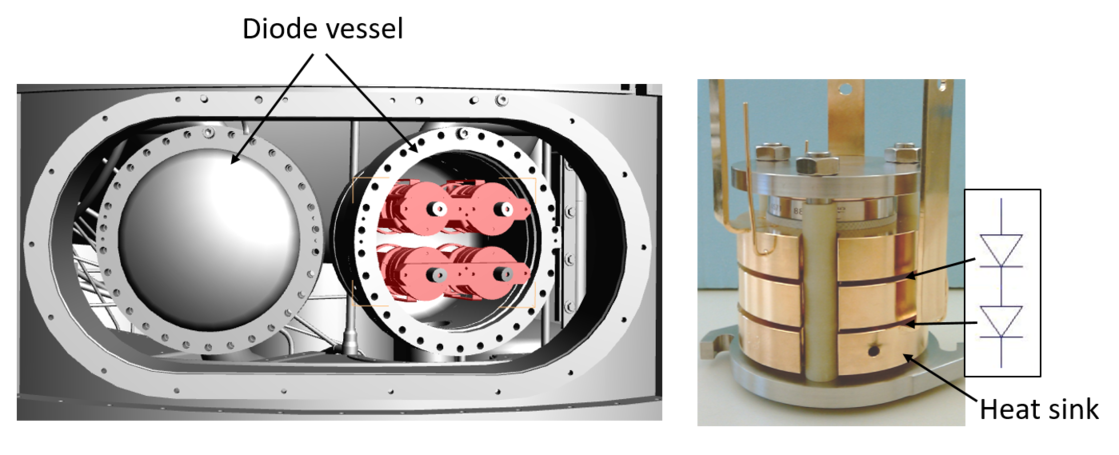}
\caption{Two diode vessels of the CPS and a photo of one diode stack with two diodes. The diode vessels (DV) are positioned at the lower part of the dome, as indicated in figure~\ref{figC0a}, where the magnetic fringe fields are negligible on the diodes.}
\label{figCd1}
\end{figure}

The cold bypass diodes are installed in two separate diode vessels in form of diode stacks (figure~\ref{figCd1}). Six modules have a diode stack with two diodes and M1 has a stack with one diode. All diode stacks have been manufactured by KIT and have been successfully cold-tested in a liquid helium environment before the installation for a current load of 10~MA$^2$s~\cite{Gil2014}. The diode stacks in the two diode vessels are easily accessible from the outside, if needed for a repair. The diode vessels are located at the lower part of the dome, where the fringe fields are negligible during standard operation.

The magnetic forces between the CPS magnets and their neighbouring magnets have been taken into account in the mechanical design. The maximum attracting magnetic force between the CPS and the first magnet of the pre-spectrometer (PS1) is 54~kN, while the force between the CPS and the new DPS magnets is 18~kN. According to the maximum magnetic forces the mechanical design of the CPS cryostat has been carried out for a slightly higher load of 65~kN, resulting in additional reinforcement ribs on the end flanges of the CPS cryostat. In addition, three special spacer bars between the CPS cryostat and the PS1 magnet had to be installed through the separation wall between the two buildings (figure~\ref{fig1}).

The CPS was designed in detail and manufactured by ASG Superconductors, S.p.A., Italy. All individual magnet modules were successfully cold-tested at the design current of 200~A before the system was delivered to KIT in summer 2015.
\subsubsection{Instrumentation for magnetic field measurement}
Two uni-axial Hall probes (Type~HHP-VP of AREPOC~s.r.o~\cite{Arepoc}) are installed on both end flanges of each module chamber for monitoring the magnetic fields. A closed-loop fluxgate type sensor IT 200-S Ultrastab~\cite{LEMit400s} manufactured by LEM$^\circledR$ is installed on the current circuit of the magnet outside of the cryostat. It has an accuracy of 0.0084$\%$. 
The water-cooled PSU of the CPS magnet was manufactured by the Bruker company with on-board current regulation electronics for a current stability of $<~\pm$~100~ppm per 8~hours. The results of the current stability of the magnet are reported in section~\ref{sec:stability}.
\subsubsection{Cooling system}
The superconducting coils of the CPS are cooled in a liquid helium bath at 4.5~K and 0.13~MPa in the same way, as described for the WGTS in section~\ref{sec:wgts}.
\subsection{Pre-Spectrometer magnets}
\label{sec:ps}
\subsubsection{Description of magnets and operation mode}
Two pre-spectromenter magnets are designed: PS1 between the entrance of the PS and the CPS and PS2 between the exit of the PS and the MS. Two 4.5~T cryogen-free conduction cooled superconducting magnets were manufactured by Cryogenics Ltd and delivered to KIT in 2003 and 2004. They were operated during many previous background studies with the pre-spectrometer (PS) ~\cite{Fraenkle2010}. The mutual attraction of the two PS magnets, separated by the 3.4 meter long pre-spectrometer, is small, but the PS1 magnet has to withstand a strong magnetic force of 54~kN, as mentioned in the previous section. The magnets have been designed with a larger margin for a maximum force of 100~kN at the beginning of the design~\cite{Gehring2004}. 

The magnets can be operated either in persistent current mode or in driven mode. However, driven mode operation is preferred with the magnets, because the field stability in driven mode with a stable power supply is better than the value of 0.2$\%$/month in persistent current mode. The result in driven mode is reported in section~\ref{sec:stability}.
\subsubsection{Instrumentation for magnetic field measurement}
Two uni-axial Hall probes of AREPOC~s.r.o are installed on the outside of both end flanges of each magnet cryostat for monitoring the magnetic field stability for long-term operation, as mentioned in section~\ref{sec:rs_mag-sensor}. In addition, a closed-loop fluxgate type sensor IT 200-S Ultrastab~\cite{LEMit400s} manufactured by LEM$^\circledR$ is installed on the current circuit of the magnet outside of the cryostat, so that it can be used in case of driven mode operation like in the CPS. 

The air-cooled PSUs of type NTS~800-5 from FuG~Elektronik~GmbH designed for a maximum current of 160~A are  measuring their currents over an 750~$\mu\Omega$ shunt. The PSUs can provide a current stability of $<~\pm$~100~ppm per 8~hours. The results of the current stability of the magnets are reported in section~\ref{sec:stability}.
\subsubsection{Cooling system}
Cooling of the cryogen-free magnets is achieved by thermal conduction with a two stage Gifford-McMahon (GM) cryocooler, Sumitomo SRDK-415D~\cite{srdk415d}. The cryocooler supplies a cooling power of 1.5~W at 4.2~K at the second stage and 35~W at 50~K at the first stage. The small Joule heating on the normal conducting part of the current leads in driven mode is the main heat source for the second stage, which is covered by the cooling power of 1.5~W. The cryocoolers require six days to cool the magnets from 293~K to 4.2~K. Seven RhFe temperature sensors are installed for monitoring the temperatures on the radiation shield, the coil, and two stages of the cold heads of each magnet.
\subsection{Main Spectrometer magnets}
\label{sec:ms}
\subsubsection{Description of magnets and operation mode}
The main spectrometer (MS) shares the fringe fields of the neighbouring superconducting magnets PS2 and the Pinch magnet, which belongs to the detector system (section~\ref{sec:detector}). The magnetic field at the analysing plane of the MS is dominated by the fringe fields of the 4.5-T PS2 magnet and the 6-T Pinch magnet. In order to optimise the fields in a range of 0.3~mT to 2~mT, there are 14, normal conducting air coils of 12.6-m-large diameter installed around the longitudinal axis of the 23.2-m-long main spectrometer. In addition, two sets of dipole coils are installed around the x- and y- axes to compensate the earth magnetic field in the MS. All 16 air coils can be individually charged by 16 power supplies for fine tuning of the magnetic field at the analysing plane. Details of the air coil systems are reported in~\cite{Glueck2013, Erhard2014}. 
\subsubsection{Instrumentation for magnetic field measurement}
The current stabilities of the 16 individual power supplies of the air-coils are monitored by the same type of DCCT as the PS magnets, IT 200-S Ultrastab~\cite{LEMit400s} manufactured by LEM$^\circledR$. 

It is very challenging to measure the magnetic fields in the MS and to determine the fields in the analysing plane with the required accuracy of 2~$\mu$T~\cite{Groh2015}. In principle, with a larger number and more precise field data points, a more accurate field analysis is possible. There are two different sensor networks installed on the surface of the main spectrometer for monitoring the magnetic field directions and the field stability; (1) 24 anisotropic magnetoresistance (AMR) sensors KMZ10B manufactured by Philips Semiconductors for a low field measurement of from 0.1~mT to a few ~mT, and (2) 14 triaxial flux gate sensors Mag-03 manufactured by Bartington Instruments~\cite{Bartington} to measure very low fields from several 10~nT to 1~mT. 

In addition, four mobile sensor units were designed by University of Fulda, Germany, using triaxial flux gate sensors FL3-1000 manufactured by Stefan Mayer Instruments. They are installed on the inner surface of four  ring frames which support the air coils. They are so-called ''Ring Magnetic field Measurement System (RMMS)'' and automatically move from a parking position along the ring frame to measure the magnetic fields at every angle of 2.5 degrees~\cite{Osipowicz2012, Letnev2018}. Another four mobile sensor units, so-called ''Vertical Magnetic field Measurement System (VMMS)'' are installed outside the air-coils on two vertical planes of east and west sides of the building. They can move vertically and horizontally to measure the remanent-and-induced magnetisation in the building.

Details about the magnetic field measurements and the field analysis are reported in~\cite{Erhard2014, Erhard2016, Erhard2017}. Further mobile flux gate sensors are still in preparation to improve magnetic field analysis.
\subsection{Detector system magnets}
\label{sec:detector}
\subsubsection{Description of magnets and operation mode}
The detector system magnets comprise the Pinch magnet and the Detector magnet that are described in~\cite{Amsbaugh2015}. The 6-T Pinch magnet provides the highest magnetic field for the experiment, while the detector magnet delivers a 3.6-T field, matching the field at the source. The magnets are designed for persistent current mode operation, having persistent switch heaters inside the liquid helium bath. They have also helium recondensing cryocooler systems like the RS- and DPS magnets, as described in section~\ref{sec:rs}. They do not have separate flanges to access cold diodes. But the small magnets can be easily accessed from outside for any repair. The magnets have been built by Cryomagnetics Ltd. Both magnets were designed for providing a maximum field of 6~T and have been successfully operated in unison. 

After successfully operating for 5 years, the first pinch magnet experienced four quenches at 5.25 T when operating with the detector magnet at 3.6T. After the quenches occurred, the Pinch magnet could no longer be charged up to 6~T when the detector magnet was charged to 3.6~T. There was evidence that the coil was moving and that this also influenced the thermal shielding. Therefore, it was replaced with a new Pinch magnet that was manufactured with two compensation coils at both ends of a main coil. The attracting force between the two magnets is about 37~kN with 6~T at the Pinch and 3.6~T at the Detector. The new Pinch magnet was successfully commissioned in 2015 and operates according to specification. 
\subsubsection{Instrumentation for magnetic field measurement}
Two uni-axial Hall probes (Type~HHP-VP of AREPOC~s.r.o~\cite{Arepoc}) are installed on the outside of both end flanges of each magnet cryostat for monitoring the magnetic field stability in long-term operation, as described in section~\ref{sec:rs_mag-sensor}. 
\subsubsection{Cooling system}
Each magnet is cooled by a Cryomech PT410 two stage pulse-tube cryocooler in a liquid helium bath with a filling volume of 0.07~m$^3$ for the Pinch and 0.08~m$^3$ for the Detector. The cryocooler supplies a cooling power of 1.0~W at 4.2~K at the second stage and 40~W at 45~K at the first stage. A heating element regulates a small over-pressure of about 4.8~kPa during liquid helium bath cooling of the coils, while the boiling helium is recondensed by the second stage cold-head of the cryocooler.
\section{Magnet safety}
\label{sec:mss}
\subsection{General safety considerations}
Operation of the strong magnetic fields in general requires to consider several safety aspects:
\begin{itemize}
\item Safety of personnel against the strong magnetic fields
\item Functionality of peripheral equipments in the vicinity of the magnets 
\item Failures of the magnets because of quench, cooling failure, power cut, etc.
\end{itemize}
This first point is particularly important for the protection of personnel with medical implants, such as pacemakers or metal implants. The safety limit for static magnetic fields is 0.5~mT according to table~B4 in the EU-directive 2013/35/EU. In addition, ferromagnetic materials have to be prevented from being carried inside an area with fields above~3~mT. 

The second point is important to ensure the uninterrupted operation of the experiment. The reliable operation of turbo-molecular pumps (TMP), vacuum pressure gauges, and electronics is often severely limited in the presence of strong magnetic fields. For instance, TMPs on the pump ports of the WGTS and DPS can heat up quite significantly due to eddy currents induced by the static fields in the fast turning rotors. The degree of heating depends on the orientation of the rotor in the magnetic field. The limits of the magnetic flux density for a large Leybold TMP MAG W2800 are about 3~mT for long-term operation and 4~mT for short term operation even with water cooling, if the rotor axis of the TMP is perpendicular to the magnetic fields~\cite{Wolf2011}. Adequate positioning and magnetic shielding are therefore important design factors. 
\begin{figure}[t]
\centering 
\includegraphics[width=150mm]{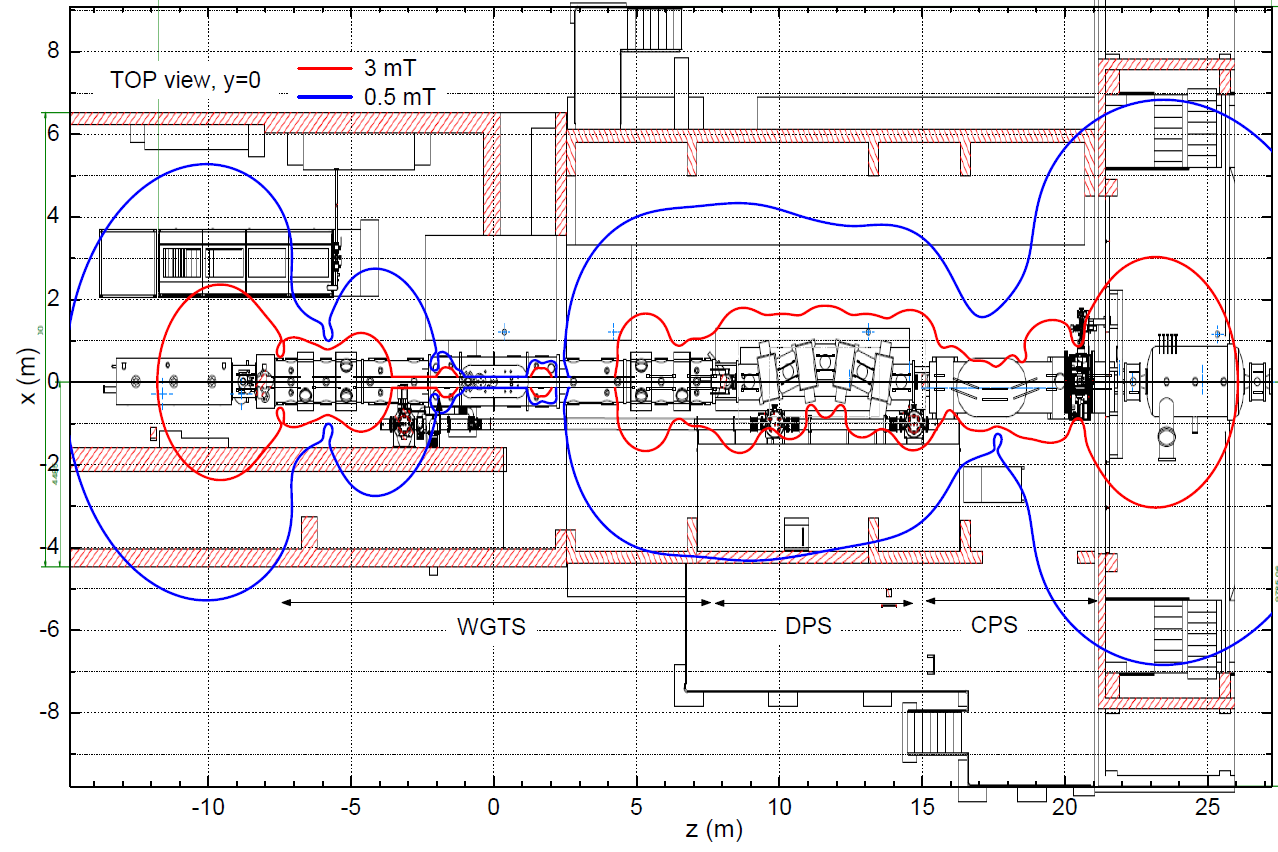}
\caption{Contour lines of the fringe fields of 3~mT and 0.5~mT in the hall of the STS. The magnetic fields are calculated for the magnets from the RS to PS1 with the following fields; RS at 4.7~T, WGTS-R and WGTS-C at 3.6~T, WGTS-F at 5~T, DPS at 5.5~T, CPS at 5.6~T, and PS1 at 4.5~T.}
\label{figStrayfield1}
\end{figure}

These two issues require calculations of the magnetic fringe fields, not only for standard operation but for magnet quenches, too. The contour lines of the fringe fields of 3~mT and 0.5~mT in the hall of the STS are shown in figure~\ref{figStrayfield1}, calculated for the design fields from the RS to the PS1 magnet, using the ''Magfield3'' code~\cite{Glueck2011b}. The calculation shows that TMPs for the isolation vacuum of the cryostat can be positioned in low field areas below 3~mT. However, 21 TMPs installed at the beam tube pumping ports and 6 small TMPs on the tritium re-cycling loops need magnetic shielding. The design of the shielding, located inside the  secondary containment, was carried out by using finite element analysis software, for example, FEMM~\cite{FEMM} for simple 2D-axisymmetric cases or Opera~FEA of COBHAM~\cite{Opera} by an external consultant\footnote{A.~Herv\'{e} with CERN.} for complicated 3D-analyses, also considering magnetic fringe field changes in case of coil quenches. As a result, ferritic steel ST37 with a thickness of 5~mm, 8~mm, and 10~mm is used for shielding of the TMPs at different positions. 

The third point is related to quenches either in its own system or an adjacent magnet. All the magnets cooled by a liquid helium bath are designed with safety valves and burst discs against over-pressure according to Pressure Equipment Directive (PED) of Europe. The design studies took into account pressure rises caused by a sudden failure of the insulation vacuum and by  magnet quenches. The safety valves of the WGTS and the CPS are set to open at 0.2~MPa over-pressure and their burst discs will rupture at 0.3~MPa over-pressure. The individual RS, DPS, Pinch, and Detector magnets are designed such that the over-pressure does not exceed 50~kPa and are therefore not subject to the PED. 

In addition, the magnets have to be protected against damage resulting from quenches, cryogenic or power failure during their long-term operation. In the next subsections we describe two different concepts of the magnet protection scheme.
\subsection{Passive protection of the RS, DPS, PS, Pinch, and Detector magnets}
In case of a quench of the small magnets (RS, DPS, PS, Pinch, detector), the stored magnet energy will be discharged through the cold bypass diodes within a typical decay time of about 1 to 1.5~seconds. The small magnets are designed so that the hot-spot temperature on the coil in case of a quench will not increase more than 150~K according to the conservative adiabatic calculation. The hot spot temperature on the pinch magnet can increase up to 227~K in case of a quench at 6~T, which is still well below the critical temperature of 350~K where the epoxy impregnation starts to weaken. A passive protection by free-wheeling cold bypass diodes is a typically well-developed protection scheme of the manufacturers. Therefore, no extra quench detection systems are implemented because of a very short quench discharge time. Nevertheless, voltage taps are accessible for diagnostic purposes and access to the protection diodes is also possible for exchange, if necessary. In the case of an emergency a quench heater may be activated, except at the PS magnets.
\subsection{Magnet protection of the WGTS and the CPS}
The two large, driven mode magnet systems, the WGTS and the CPS, require magnet safety systems (MSS) with an external dumping unit in addition to the cold bypass diodes for each current circuit. Furthermore, quench heater activation on the quenched module is foreseen to homogeneously distribute heat and to reduce the hot-spot temperature. The reason for the MSS with external dumping unit is that the slow discharge time $\tau_{sd}$ of the large magnet systems in driven mode is several tens or even 100 times longer in case of a coil quench than the quench discharge time of one single magnet. The slow discharge time $\tau_{sd}$ depends on the number of quenched coils. The maximum slow discharge time $\tau_{sd}$ of the large magnet systems occurs in case of a degradation of the bus bars, resulting in their over-heating without a coil quench, as summarized in table~\ref{tab3}. Thereby, high current loads ``MIITs'' are expected on the resistive components according to  
\begin{equation}
\label{eqMIIT}
MIITs~=~\int{I(t)^{2}dt}.
\end{equation}
With the exponentially decreasing current $I(t) = I_{0}\exp{(-t/\tau_{sd})}$ with a decay time $\tau_{sd}~=~L/R_d$, depending on the coil inductance $L$ (table~\ref{tab1}) and external dump resistor $R_d$,  the integration of eq.~(\ref{eqMIIT}) simply results in
\begin{equation}
\label{eqMIIT2}
MIITs~=~I^{2}_{0}\dfrac{\tau_{sd}}{2}.
\end{equation}
\begin{figure}[t]
\centering 
\includegraphics[width=70mm]{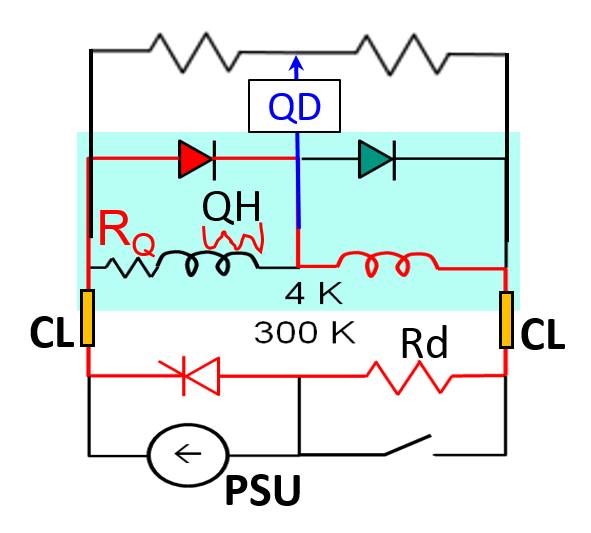}
\caption{Scheme of driven mode operation with an external dumping unit. QD: quench detector, QH: quench heater, $R_Q$: resistance at quench, $R_d$: external dump resistor. In case of a quench the current flows through current leads (CL), the external dumping unit, the cold diode on the quenched coil, and the unquenched coil of the magnet.}
\label{figDM}
\end{figure}

Assuming a constant R$_{d}$ and a time delay of 1.5~s for the circuit breakers to open, the calculated values of MIITs for a slow discharge without a coil quench are summarized for each magnet current circuit of the WGTS and the CPS in table~\ref{tab3}. The high values of MIITs mean higher Joule heat generation in the cold bypass diodes and in the external dumping unit; for example, bus bars, external diodes, and dump resistor (figure~\ref{figDM}). Therefore, their dimensions have to be properly chosen for the maximum value of MIITs of the system. For instance, the heat sinks of the cold bypass diode stacks of the CPS are designed for a MIITs of 10~MA$^2$s, while the diode stacks of the WGTS are designed for a MIITs of 4~MA$^2$s. All diode stacks successfully passed the current endurance tests at 4.2~K before the assembly~\cite{Gil2014}, having the maximum temperatures on the diodes below 250~K. The 21~mm$^2$ bus bars of the CPS are sufficiently large to withstand the maximum MIITs of the CPS while the bus bars of the WGTS are 5~mm$^2$ small, which could be critical in a conservative adiabatic calculation without cooling power. The temperature increase of the bus bars for the maximum MIITs can be calculated by a simple adiabatic heat balance relation with the heat capacity of the copper bus bar according to~\cite{Wilson1983}
\begin{equation}
\label{eqT1}
\int{I(t)^{2}dt} = A^{2}_{cs}\int^{T_{max}}_{T_{i}}{\dfrac{\gamma•C(T)}{\rho(T,B)}dT},
\end{equation}
where $A_{cs}$, $\gamma$, $C(T)$, and $\rho(T)$ are the cross-section area, the mass density, the specific heat capacity, and the specific resistivity of the bus bar with respect to temperature ($T$) and $T_i$ is the initial temperature and $T_{max}$ is the maximum temperature. $\rho(T)$ slightly depends also on the magnetic flux density ($B$) because of the magnetoresistance effect. The temperature of the small bus bars calculated for the MIITs could increase up to 580~K where the soft soldering connections will start to melt. Therefore, such high temperatures have to be avoided by a proper magnetic safety system, although this is a conservative calculation, because the discharge time at a coil quench will be reduced due to inductance reduction by subtracting the quenched coil. Thus the MIITs values of the WGTS-C will be below 2.7~MA$^2$s and the temperature of the bus bars can increase only up to around 270~K. If there is a risk of overheating the bus bars or the current leads, they have to be protected by additionally triggering the activation of the quench heaters to quickly discharge the affected magnet. 
\begin{table}[t]
\caption{Main parameters of the external dumping units of the WGTS and the CPS. R$_d$: external dump resistor, $\tau_{sd}$: time constant for slow discharge without quench heater activation. MIITs~(unit: MA$^2$s) according to eq.~(\ref{eqMIIT}) with a delay of 1.5~s for breaker opening.}
\label{tab3}
\centering
\vspace*{1ex}
\begin{tabular}{|l|c|c|c|c|}
\hline
Circuits&WGTS-R&WGTS-C&WGTS-F&CPS\\
\hline
R$_d$~($\Omega$)&0.413&0.413&0.497&0.45\\
$\tau_{sd}$~(s)& 66&81&150&382\\
MIITs &3.3&4.0&3.3&7.7\\
\hline
\end{tabular}
\end{table}

Furthermore, because of non-negligible inductive coupling due to short distances between the magnet modules, the distinct quench detection is rather complicated, especially for the WGTS and the CPS. The mutual magnetic coupling coefficients $k_{ij}$ between \textit{i-th} and \textit{j-th} coils is defined by
\begin{equation}
\label{eqkij}
k_{ij} = M_{ij}/\sqrt{L_{i}L_{j}},
\end{equation}
where $M_{ij}$ is the mutual inductance between \textit{i-th} and \textit{j-th} coils, and $L_i$ and $L_j$ are the self-inductances of the coils. The inductance matrix of the magnets was analytically calculated in MathCAD~\cite{MathCAD} according to the algorithm reported in~\cite{Fawzi1978}. For the inductance matrix of the WGTS magnets, a total of 30 coil segments including the RS and DPS-M1 were calculated. For the inductance matrix of the CPS magnets, a total of 17 coil segments including the DPS-M5 and PS1 were calculated. The selected coupling coefficients $k_{ij}$ for the WGTS and the CPS are summarized in table~\ref{tab4}. The coupling coefficients of the WGTS show that the compensation coils of the three long modules are more stronger coupled with each other because of the short distances, as expected. The coupling coefficients of the CPS show that the outer winding segments of the modules are slightly stronger coupled than the inner windings because of the larger diameter, as expected. The coupling coefficient between L6B and L7B is slightly smaller than the others because of the larger separation distance by the pumping port. 
\begin{table}[t]
\caption{Inductive coupling coefficients $k_{ij}$ between neighbouring coil modules of the WGTS and the CPS. \textit{i-th} coil and \textit{j-th} coil are selected for the case with the highest value of $k_{ij}$. Typically they are the adjacent compensation coils from each other. See figure~\ref{figWe1} and \ref{figCe1} for the coil labels.}
\label{tab4}
\centering
\vspace*{1ex}
\begin{tabular}{|l|c|c||l|c|c|}
\hline
\multicolumn{3}{|c||}{WGTS}&\multicolumn{3}{|c|}{CPS}\\
\hline
\textit{i-th} coil & \textit{j-th} coil & $k_{ij}$ & \textit{i-th} coil & \textit{j-th} coil & $k_{ij}$\\
\hline
RS & L5A &0.0090 & DPS-M5 & L1 &0.0057\\
L5B & L4A & 0.0126 &L1 & L2B & 0.0223\\
L4B& L1A & 0.0127 &L2B& L3B & 0.0232\\
L1B & L2A & 0.2742&L3B & L4B & 0.0233\\
L2B & L3A & 0.2742&L4B & L5B & 0.0209\\
L3B & L7A & 0.0171&L5B & L6B & 0.0352\\
L7B & L6A & 0.0214&L6B & L7B & 0.0148\\
L6B & DPS-M1 & 0.0051&L7B & PS1 & 0.0117\\
\hline
\end{tabular}
\end{table}

Because the coupling coefficients of the WGTS and the CPS magnets are relatively small, a coil quench is unlikely to induce a cascaded quench in its neighbouring coils, considering the safety margin of the conductor of more than 20$\%$ against quench. However, the inductive coupling is not negligible for quench detection. If no cascaded quench is expected and a fast discharge is not necessary for other reasons, then it is better to activate the quench heaters on the quenched coil module only, leaving all other un-quenched coils undisturbed. This will help to reduce additional thermal stress by the quench heaters, because quench heaters are foreseen to homogeneously distribute the locally generated heat to the other areas of the already quenched coil module. Without distinct quench detection an effective activation of the quench heaters is not possible for the inductively coupled magnets. Therefore, the influence of the inductive couplings on the quench detection has been studied by electrical quench simulations and a distinct quench detection method has been developed for the inductively coupled KATRIN magnets.
 \subsubsection{Distinct quench detection}
The concept of the distinct quench detection method applied for the CPS and the WGTS magnet systems is presented in~\cite{Gil2016} and the first quench detection performances are reported for the CPS in~\cite{Gil2017} and for the WGTS in~\cite{Gil2017b}. In this section we briefly present the concept of the distinct quench detection method, using a programmable logic controller (PLC) in combination with a global detector (GD) and one or more quench detectors. 
\begin{figure}[t]
\centering 
\includegraphics[width=150mm]{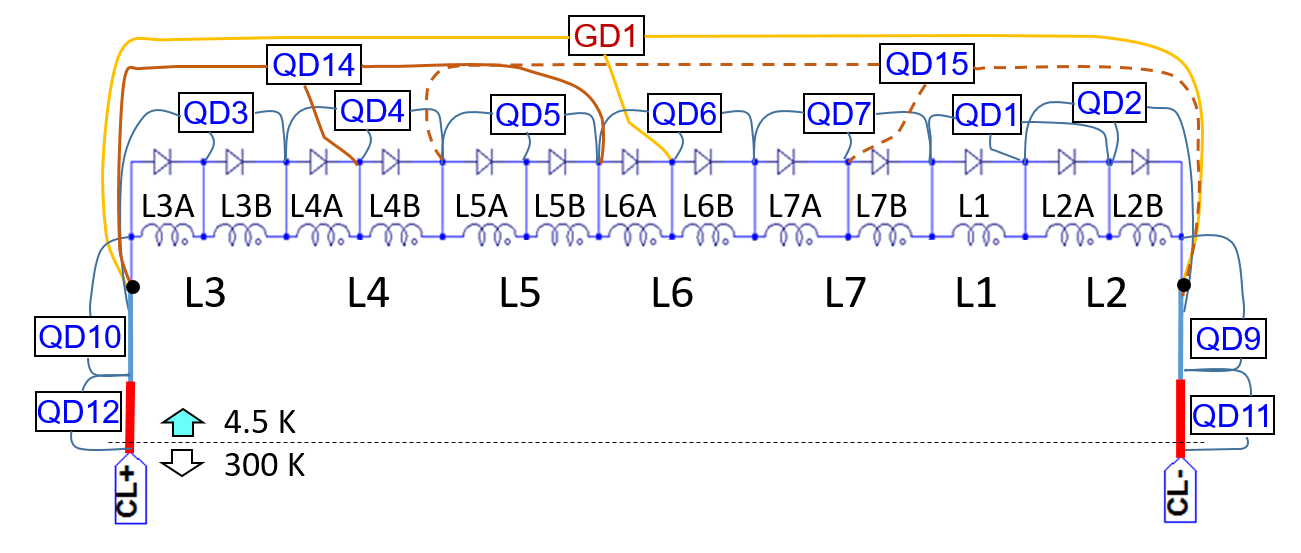}
\caption{Scheme of the quench detectors (QD) of the CPS magnets. ``Ln'' denotes \textit{n-th} module. ``A'' denotes the inner winding segment of ``Ln'' and ``B'' the outer winding segment of ``Ln''. ``QDn'' denotes \textit{n-th} quench detector (QD). ``GD1'' indicates a global quench detector. The detectors from QD9 to QD12 are detectors on the current leads and bus bars. QD14 and QD15 are additional redundant detectors which are not mandatory. The current leads (CL) are connected to the external units, as shown in figure~\ref{figWe1}d.}
\label{figCqds1}
\end{figure}
\begin{figure}
\centering 
\includegraphics[width=140mm]{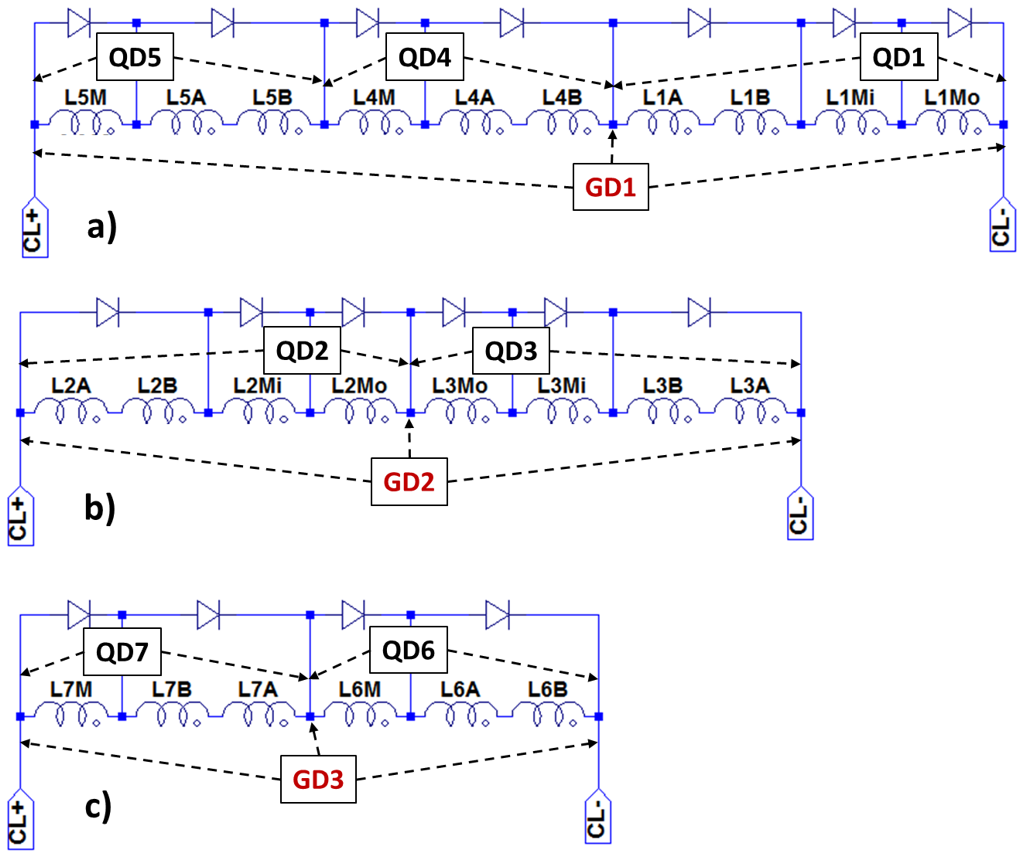}
\caption{Scheme of the quench detectors (QD) of the WGTS magnets. ``Ln'' denotes \textit{n-th} module. ``A'' denotes the inner winding segment of ``Ln'' and ``B'' the outer winding segment of ``Ln''. ``QDn'' denotes \textit{n-th} quench detector (QD). Three global quench detector (GD1 to GD3) are installed for each main circuit. The detectors on the bus bars and current leads and on the dipole coil pairs are not shown for simplicity. The current leads (CL) are connected to the external units, as shown in figure~\ref{figWe1}d.}
\label{figWqds1}
\end{figure}
\begin{figure}[t]
\centering 
\includegraphics[width=140mm]{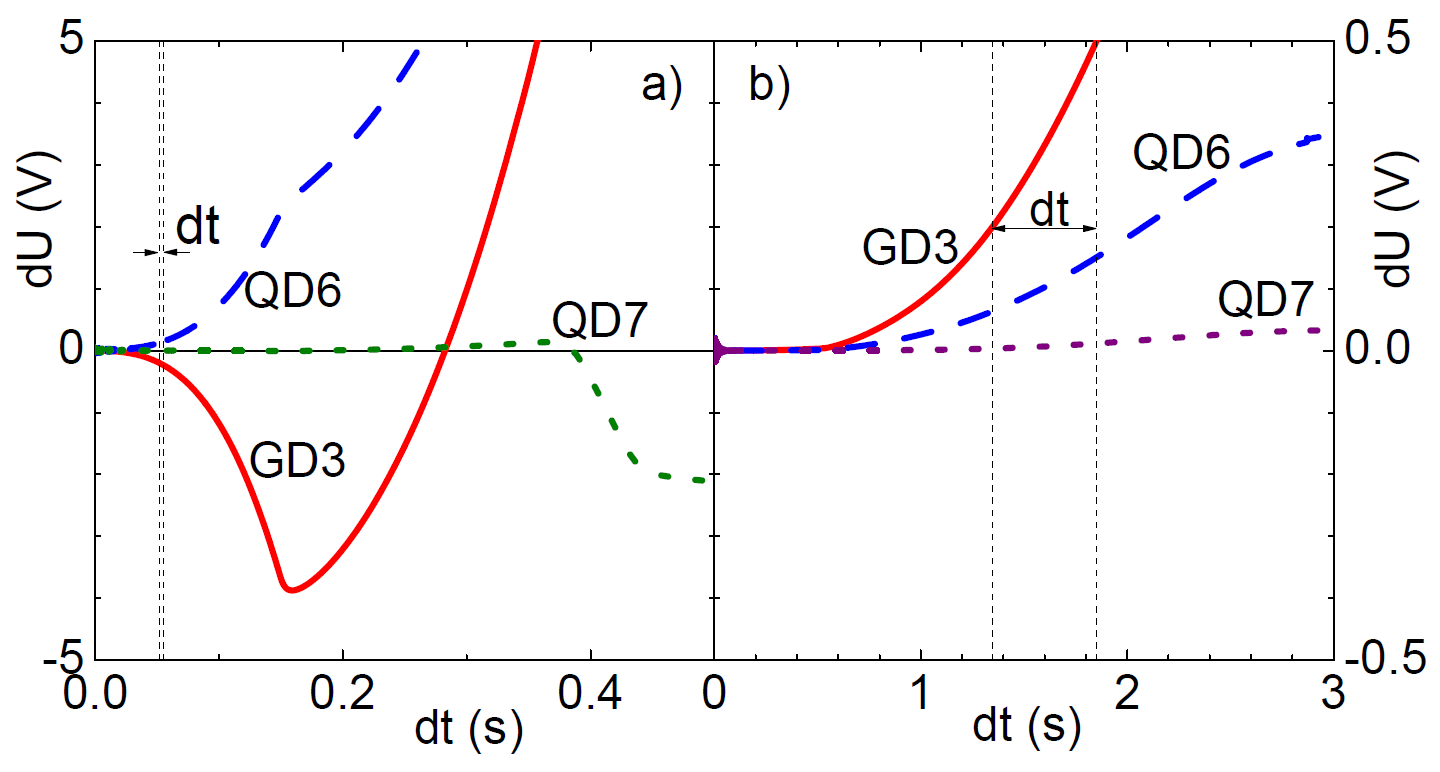}
\caption{Example of an electrical quench simulation for a quench in M6 (a) and for a quench in DPS-M1 (b). ''dU" indicates the unbalanced voltage drop of a detector. The time interval between GD3 and GD6 is below 10~ms for a) and about 500~ms for b).}
\label{figQab}
\end{figure}

At first, the conventional quench detectors (QD) based on the bridge method~\cite{Wilson1983} have been installed to monitor an unbalanced voltage of the two half coil segments in each solenoid module. Figure~\ref{figDM} shows one magnet module in driven mode with an external dumping unit. The external dumping will be activated by opening the circuit breakers either for slow discharge or at a quench. The quench detector (QD) has to be balanced by setting a gain factor $G$ during first ramping to compensate the different voltages at each half coil because of $dI/dt$ of the current circuit. The voltage drops $U_1$ and $U_2$ on coil~1 ($L_1$) and coil~2 ($L_2$) can be described by
\begin{eqnarray}
\label{eqLRMa}
U_{1} &=& L_{1}\dfrac{dI_1}{dt} + I_{1}(t) R_{Q}(t) + M_{12}\dfrac{dI_2}{dt},\\
\label{eqLRMb}
U_{2} &=& L_{2}\dfrac{dI_2}{dt} + M_{21}\dfrac{dI_1}{dt},
\end{eqnarray}
where $R_Q$ is the quench resistance at a quench of coil~1 and $M_{12}~=~M_{21}$ is the mutual inductance. The bridge voltage of the QD, dU~=~$G_{2}U_{1}~-~G_{1}U_{2}$~=~0, is balanced by adjusting a gain factor G$_{1}$ for $U_{1}$ relative to a gain factor G$_{2}$ for $U_{2}$ during ramping. With G$_{2}~\equiv$~1, $U_{1}~=~G_{1}U_{2}$ for dU~=~0. The gain factor $G_{1}~=~(L_{1}+M_{12})/(L_{2}+M_{21})$ is determined by $dI_1/dt$~=~$dI_2/dt$ at $R_Q$~=~0. After the balancing with the gain factor the detector is ready for detection of unbalanced voltages based on the resistive transition at a quench (R$_Q$).

The voltage drop $U_i$ on the \textit{i-th} coil, which is inductively coupled with all other coils in different current circuits, will generally behave according to 
\begin{eqnarray}
\label{eqLRM1}
U_{i}~&=&~L_{i}\dfrac{dI_i}{dt} + I_{i}(t) R_{Q,i}(t) + \sum^{j}{M_{ij}\dfrac{dI_j}{dt}},
\end{eqnarray}
where $M_{ij}$ is the mutual inductance between \textit{i-th} coil and \textit{j-th} coil. The gain factor $G_{ij}$ between \textit{i-th} and \textit{j-th} coils against all other coils can be balanced by adjusting $U_{i}$ = $G_{ij}U_{j}$ during a synchronized ramping of all coils, without quench during the balancing. $R_{Q,i}$ is considered only for a quench of the \textit{i-th} coil.

The arrangements of the QDs for each solenoid module are shown for the CPS in figure~\ref{figCqds1} and for the WGTS in figure~\ref{figWqds1}. The minimum number of the detectors for reliable quench detection in these coils are 8 detectors for the CPS (one GD for one circuit and 7 QDs for 7 modules) and 10 detectors for the WGTS (three GDs for three circuits and 7 QDs for 7 modules).  

Furthermore, a PLC is needed to validate a quench according to proper logical rules and to avoid spurious detection caused by the inductive couplings. 30 coil segments for the WGTS and 17 coils for the CPS including their neighbours were taken into account for the study of the inductive couplings of the quench detection, which has been carried out by electrical quench simulations with LTspice~IV of LINEAR TECHNOLOGY~\cite{LTspice}.

In this section we focus on a method whereby a quench in a coil can be distinguished from the quenches in other neighbouring coils. An example of the electrical quench simulation is presented for a quench in M6 of the WGTS (figure~\ref{figQab}a) and for a quench in DPS-M1 (figure~\ref{figQab}b). For both quenches, the unbalanced voltages of the detectors GD3 and QD6 are higher than the detection threshold values of 200~mV for GD3 and 150~mV for QD6. Therefore, it is not possible to distinguish the two quenches just by the quench detectors without proper logical rules. However, the quench propagation in M6 is very fast within a few 100~ms, while the unbalanced voltages at the detectors of M6 for the quench in DPS-M1 increases very slowly because of the weak inductive coupling. The time interval between GD3 and GD6 is below 10~ms for the quench in M6 (figure~\ref{figQab}a), while it is about 500~ms for the quench in DPS-M1 (figure~\ref{figQab}b). This simulation indicates a first logical rule for a distinct quench validation, combining one GD and one QD with a proper time interval ($t_{QV}$). Table~\ref{tab4b} shows an example of two logical rules for these cases in order to distinguish quench in module M6 from a quench in the DPS-M1. Rule no.~1 detects a quench in M6 with a validation time interval of 80~ms between QD6 and GD3, while rule no.~2 detects a quench in DPS-M1 with a longer time interval between QD6 and GD3. After several simulations for different quench cases, the logical rules with proper parameters have been defined for an unambiguous quench signal of each solenoid module. Details of the main logical rules and some experimental results for the distinct quench detection are reported in~\cite{Gil2016, Gil2017, Gil2017b}. Even symmetric quenches which are usually not possible with just one QD can be detected in combination with one GD and two neighbouring QDs, too~\cite{Gil2016}.
\begin{table}
\caption{Example of two logical rules for distinct quench detection for module M6 of the WGTS magnet. See figure~{\ref{figWqds1}} for the position of M6, QD6 and GD3. `QV' denotes Quench Validation by QD6 and GD3 within a time interval (t$_{QV}$) between these detectors' events.}
\label{tab4b}
\centering
\vspace*{1ex}
\begin{tabular}{|l|c|c|c|c|}
\hline
Quench & \multicolumn{2}{c|}{QD combination}&t$_{QV}$&simulation\\
\hline
M6(=L6)&QD6&GD3&t$_{QV}$ $\leq$ 80~ms&figure~\ref{figQab}a\\
DPS-M1&QD6&GD3&80~ms $<$ t$_{QV}$ $\leq$ 1.5~s&figure~\ref{figQab}b\\
\hline
\end{tabular}
\end{table}
\subsubsection{Realised magnet safety systems}
The programmable logic controller (PLC), Simatic S7 from Siemens AG, manages relevant interlocks of the magnet safety system (MSS) for magnet protection according to the logical rules as follows:
\begin{figure}[t]
\centering 
\includegraphics[width=120mm]{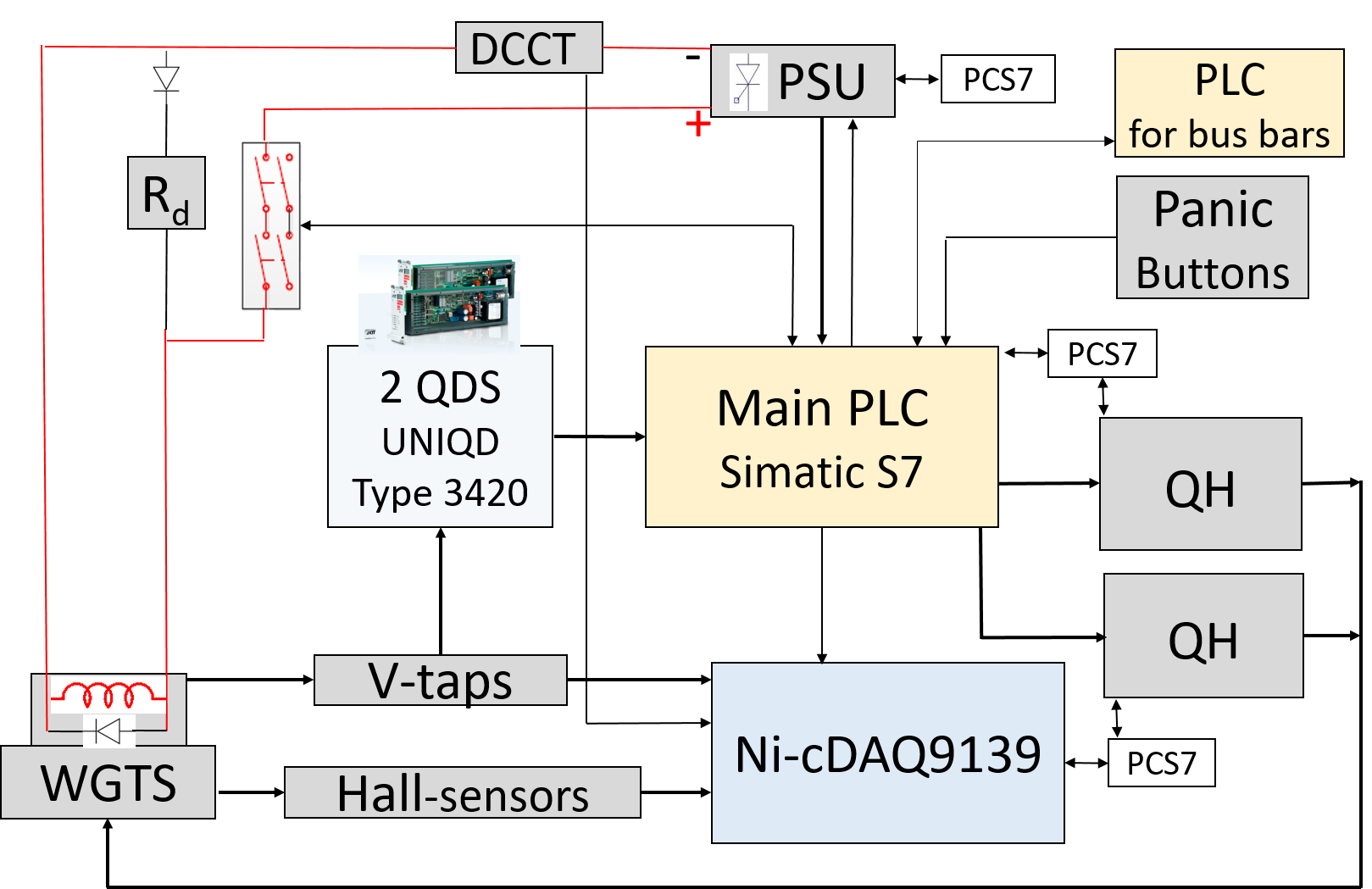}
\caption{Scheme of the MSS of the WGTS magnet system. The magnets are connected with the external dumping unit and the power supply unit (PSU), shown here for one of the three circuits only for simplicity. The main programmable logic controller (PLC) is monitoring the signals of the 41 detectors in the two QDS, PSU, Panic buttons, and those of the cryogenics via PCS7. It also triggers quench heaters (QH) and a DAQ system. A redundant PLC for the bus bars protection also communicates with the main PLC. DCCT stands for DC current transducer.}
\label{figWmms1}
\end{figure}
 
Once a quench is detected by the detectors of the quench detection system (QDS), UNIQD Type 3420 from KIT~\cite{IPE},\footnote{The universal quench detection system UNIQD was first developed for the superconducting magnet system of the stellarator Wendelstein 7-X~\cite{Birus2007}.} the detector signals are analysed and confirmed by the PLC of the MSS. A flow diagram of the MSS of the WGTS is shown in figure~\ref{figWmms1}.

After a quench has been confirmed, the PLC has to trigger interlocks for magnet protection according to the pre-defined protection rules~\cite{Gil2016, Gil2017b}. Firstly, the MSS has to open the breakers to discharge the stored magnetic energy through the external dumping units for each coil quench and for other major failures (e.g. cryogenic failures, power failure, etc.).

In addition, the PLC has to activate the quench heaters on the quenched module (figure~\ref{figQab}a) in order to homogeneously distribute the heat generation in the quenched module. If there is no quench in the magnet itself, but in one of the neighbouring magnets (for instance, figure~\ref{figQab}b), the PLC only initiates a slow discharge of the magnet by opening the breakers without quench heater activation. This will reduce liquid helium loss and recovery time. If a distinct quench detection were not possible, the usual course of action would be the activation of all quench heaters on all coils of the magnet system for a fast discharge. The disadvantage of such a global discharge would be the risk of an over-pressure followed by the rupture of a bursting disc of the large liquid helium chamber, causing a loss of large amounts of helium, more thermal stress for the coils, high costs, and long downtime of the system. Thus, the MSS with the distinct quench detection allows us to effectively protect the magnets and save operation costs and time.
\section{First commissioning results}
\label{sec:commissioning}
\subsection{Simultaneous operation of all magnets}
Operational runs of the magnets are summarized in table~\ref{tab6aa}. During the commissioning phase Run~$\sharp$0, all small magnets and the WGTS magnets were successfully tested up to their design currents. One module (M7 of figure~\ref{figCe1}) of the CPS magnets had a training quench at 194~A (97$\%$ of its design current) during the first energizing. The quench was detected by its magnet safety system~\cite{Gil2017}. The magnet has been re-energised up to 90$\%$ of the design current for the magnetic force test with the PS1 magnet in order to avoid further quench risk.

DPS-M1 had a quench during the synchronized ramping of the KATRIN magnets (Run~$\sharp$1) just before reaching 50$\%$ of its design current because of a heat input caused by a small leak in the indium seal of the diode flange of the helium chamber. This quench was detected by the MSS of the WGTS, resulting in a safe slow discharge of the WGTS magnets. More details on the quench detection performance for this quench is reported in~\cite{Gil2017b}. 
\begin{table}[t] 
\caption{Summary of magnet operation runs. Run $\sharp$0: single magnet operation up to 100$\%$ of $I_d$ for commissioning. Run $\sharp$1: ''first-light" test at 50$\%$ of $I_d$. Run $\sharp$2: ''first-light-plus" test at 20($\pm$5)$\%$ of $I_d$. Run $\sharp$3: ''krypton calibration" test at 70$\%$ of $I_d$. ''y" means "operated" and ''n" for ''not-operated''. ''Q.'' indicates ''Quench at a given percentage of its design current''. $t_{run}$: time of run duration in days.}
\label{tab6aa}
\vspace*{1ex}
\begin{tabular}{|c|*{12}{c|}}
\hline
Run&$t_{run}$&RS&WGTS&\multicolumn{5}{c|}{DPS}&CPS&PS&\multicolumn{2}{c|}{Detector}\\
\cline{5-9} \cline{12-13}
~&~&~&~&M1&M2&M3&M4&M5&~&~&PCH&DET\\
\hline
$\sharp$0&$\geq$~0.3&y&y&y&y&y&y&y&Q.~97$\%$&y&y&y\\
$\sharp$1&4&y&y&Q.~48$\%$&y&y&y&y&y&y&y&y\\
$\sharp$2&14&y&y&y&y&y&y&y&y&y&y&y\\
$\sharp$3&$\geq$~10&n&y/n&y/n&y/n&y/n&y/n&y/n&y&y&y&y\\
\hline
\end{tabular}
\end{table}

The `first-light' test with Run $\sharp$1 successfully demonstrated for the first time the transport of electrons through the entire 70-m-long beam line from the Rear Section to the detector at a reduced field of 50$\%$~\cite{CERN2016, KATRIN2018a}. For this test the magnets were operated in non-standard field configurations because of the quench of DPS-M1. After the leak-tightness of the diode flange of DPS-M1 had been repaired, all the magnets ran at the reduced fields (20~$\pm$5$\%$) for about two weeks for further tests of the beam alignment and other background tests in Run~$\sharp$2.
 
After the installation of further components (Condensed Krypton Source and other inserts) inside the beam line, the KATRIN magnets were operated for the krypton measurements with the high voltage of the spectrometer section energized. For this first krypton measurement, Run~$\sharp$3, the RS magnet was not needed. The magnets of the SDS were charged before the STS magnets. The STS magnets were charged to 70$\%$ of their design fields in order to operate them at a safe level without the risk of a quench. The magnets were quasi-synchronously ramped to reduce the mutual inductive influence between them. The fields were set at 70$\%$ of their design values in order to operate the magnets at a safe level without a quench risk of the complex systems. The magnets were operated for about three weeks at 70$\%$ of the design currents. The magnets of the WGTS and the DPS were no longer needed after 10 days, since the tests continued with the condensed krypton source in the CPS~\cite{KATRIN2018a}.

The three-week operation of Run~$\sharp$3 was a first test of the long-term stability of the system at 70$\%$ of the design field for the KATRIN neutrino mass measurements, which  will be operated in 60 days intervals. This test did not include the tritium circuits, which were not yet connected to the WGTS at the time of the measurements.
\subsection{Experience with instrumentation}
\label{sec:sensors}
The sensors for the magnetic field stability are described for each magnet in section~\ref{sec:ka-mag}. There was no significant issue with the instrumentation because redundant sensors are installed. Some experiences with the sensors up to now are briefly summarized below. 
\begin{itemize}
\item{Hall sensors}: Number of failed Hall sensors is one of 14 in the WGTS and four of 14 in the CPS. They are installed inside the magnet chambers. But there is still at least one Hall sensor on each magnet module in operation. 
\item{Voltage taps}: Number of failed voltage taps is one of 55 in the WGTS and also one of 42 in the CPS. There is a spare voltage tap for the failed position. 
\item{Persistent switch heater}: One persistent switch heater of DPS-M1 has failed. But it could be re-installed through the diode bar. 
\item{DC current transducer DCCT}: DCCT sensors are installed through the cables that are connected from the PSU to the current leads of the magnets with driven mode. The secondary measurement circuit of the DCCT sensors need to be installed with a clear grounding point to avoid any unwanted influence upon each other. Some DCCT sensors are installed in combination with Knick isolation amplifier Type VarioTrans~P27000 for the magnet safety systems of the WGTS and the CPS. Their current fluctuations are higher than the one of the PSU because of the gain error (about 0.08$\%$) of the Knick isolation amplifier.\footnote{Technical data, https://www.knick-international.com/export/media/1302.pdf} The fluctuation of the current sensor in the stable PSU of the systems is within the specification of below 0.01$\%$. The current stabilities of the stable PSUs of the WGTS and the CPS are summarized in the next section.
\end{itemize}
\subsection{Magnetic field stability}
\label{sec:stability}
The magnetic field stability of the superconducting magnets operating in the persistent current mode is checked by measuring the magnetic field drift with a NMR probe with an accuracy of $\pm$~5 ppm per day, once the magnets are switched into the persistent current mode. The magnetic field stability of the magnets with driven mode is checked by the current stability of their power supplies, as described in sections~\ref{sec:wgts} and \ref{sec:cps}. The magnetic field stability at the analysing plane of the MS is checked by the precise flux gate sensors mounted on the surface of the MS in section~\ref{sec:ms}.   
\subsubsection{Single magnets with persistent current mode}
\begin{figure}[t]
\centering 
\includegraphics[width=80mm]{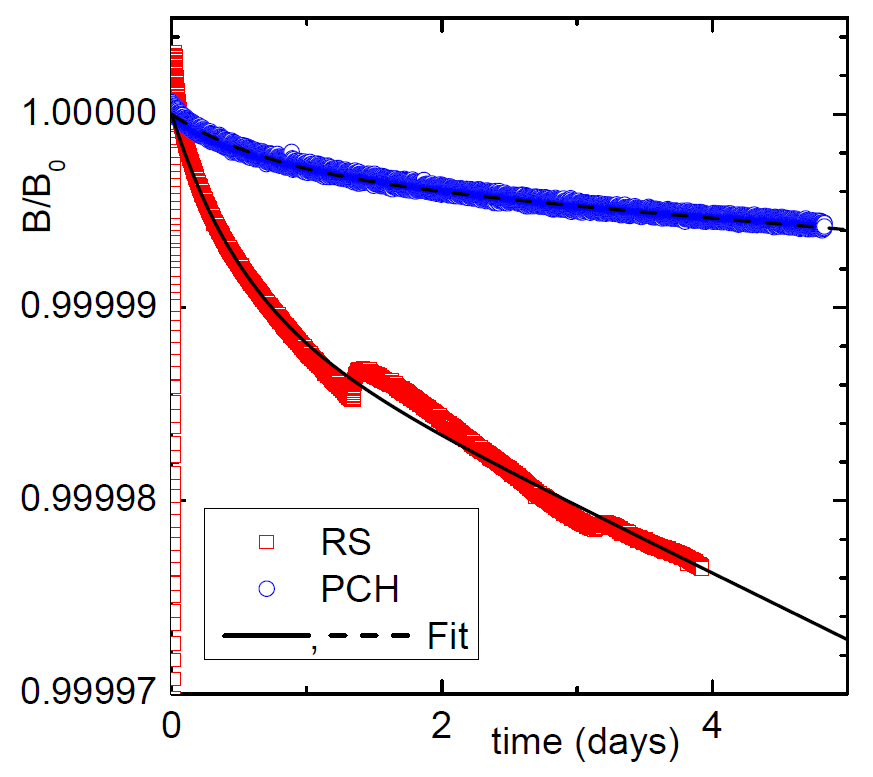}
\caption{Drifts of the normalised magnetic flux densities $B/B_0$ measured for the Pinch and the RS magnets in the persistent current mode by a NMR probe. The relative uncertainty of the NMR probe of Metrolab PT2025 is below $\pm~5.0\times10^{-8}$. The data was fitted with eq.~(\ref{eqDrift1}).}
\label{figPch1}
\end{figure}

During commissioning, the magnetic field drifts of the single magnets in the persistent current mode (PM) were checked at the manufacturer. The results are within the specified values of 0.1$\%$/month for the DPS and RS magnets and 0.01$\%$/month for the pinch and detector magnets. The magnetic field drifts $\Delta B/B_{0}$ of these magnets are summarized in table~\ref{tab6}~(bottom). 

The measurements of the magnetic field drifts of the RS, PCH and DET magnets were also performed in the persistent mode over a period of several days at KIT. Figure~\ref{figPch1} shows the drift of the magnetic fields of the PCH and the RS magnets which are normalised to their initial fields. 

The exponential drift of the magnetic flux density of a simple superconducting coil can be described by a time constant $\tau~=~L_s/R_j$ with self-inductance $L_s$ and a total joint resistance $R_j$ of the coil. However, the drift behaviour of a superconducting magnet cannot be described by one decay time only because of the complex behaviour of the currents in the multi-filamentary superconducting wires. The dominant ''transport current" of the superconducting wires slightly decreases because of the inductive coupling with the ''screening current". The screening current can be induced on the surface of the multi-filamentary superconductor of the coil-winding during the current ramping ($dI/dt$)~\cite{Cesnak1977}. Taking into account two decay times $\tau_1$ for the transport current and $\tau_2$ for the screening current and a coupling coefficient ($\alpha$) between these two currents, the drift behaviour of the magnetic field over time can be qualitatively described by a simplified function, as already applied for other measurements in~\cite{Gil2012, Amsbaugh2015}: 
\begin{equation}
\label{eqDrift1}
B(t) = B_{0}\lbrace (1+\alpha) e^{\frac{-t}{\tau_{1}}} - \alpha e^{\frac{-t}{\tau_{2}}} \rbrace,
\end{equation}
where $B_{0}$ is the initial magnetic flux dentisty. 

The fit has been performed for the data from 2~hours because of the higher drift at the beginning of the persistent mode. The fit parameters are summarised in table~\ref{tab6a}. The drift calculated with eq.~(\ref{eqDrift1}) is 38~ppm in the PCH after 60~days and 217~ppm in the RS~\cite{Gil2018a}, which are roughly ten times better than the specifications for both cases. 

Small field-recovery effects were observed in the NMR measurement of the RS after about one day and three days, which were also observed in another large magnet system~\cite{Gil2012}. The complex behaviour of the small field-recovery effect cannot be explained by the simple function. It could be related to irregular redistribution of the screening currents because of flux jumping in the multi-filamentary superconductor during the field decay. However, this effect is negligible, since it is below 8~ppm and is decreasing with time. 
\begin{table}[t]
\caption{Fit parameters according to eq.~(\ref{eqDrift1}).}
\label{tab6a}
\vspace*{1ex}
\centering
\begin{tabular}{|l|c|c|}
\hline
Magnet & PCH & RS\\
\hline
$B_0$~(T) &6.0099215~$\pm$~1.68$\cdot$10$^{-8}$&4.95572~$\pm$~1.0$\cdot$10$^{-5}$\\
$\alpha$ &-3.0433$\cdot$10$^{-6}$~$\pm$~4.3$\cdot$10$^{-9}$ &-9.9577$\cdot$10$^{-6}$~$\pm$~0.0\\
$\tau_{1}$~(s) &1.47228$\cdot$10$^{11}$~$\pm$~2.99$\cdot$10$^{8}$ &2.50596$\cdot$10$^{10}$~$\pm$~1.52$\cdot$10$^{7}$ \\
$\tau_{2}$~(s) &6.5827$\cdot$10$^{4}$~$\pm$~2.0$\cdot$10$^{2}$ &4.6447$\cdot$10$^{4}$~$\pm$~77.8 \\
Adj.-R$^2$ &0.9954 &0.9939\\
\hline
\end{tabular}
\end{table}
\begin{table}[t]
\caption{Instability of the magnetic field of each magnet group at 70$\%$ of the design fields. First table is for the DM magnets with PSU. `dt' indicates the time interval used for the calculation of $I_{av}$. The data from day~1 to day~7 were taken for the calculation of $I_{av}$ for the WGTS, because their current has been reduced to 50$\%$ at day~7. The second table is for the PM magnets. $\Delta B/B_{0}$ is calculated for 30~days from the NMR measurements.}
\label{tab6}
\vspace*{1ex}
\begin{tabular}{|l|c|c|c|c|c|c|}
\hline
Magnet with DM & WGTS-R & WGTS-C & WGTS-F & CPS & PS1 &PS2\\
\hline
$I_{av}$ (A) & 216.586 &215.877& 145.924 & 140.023 &109.435&108.762\\
$\sigma$ (A) & 0.0035 &0.0028& 0.0026 & 0.0016&0.0056&0.0207\\
$\Delta I/I_{av}$ ($\%$) & 0.0016 & 0.0013& 0.0018 &0.0012&0.019&0.0051\\
I-drift ($\%$/30d) & -0.019 & -0.017& -0.020 &-0.00014&-0.007&-0.046\\
dt (h) & 148.5 &148.5 &148.5 &475 &477.8 &480\\
\hline
\end{tabular}
\begin{tabular}{|l|c|c|c|c|c|c|c|c|}
\hline
Magnet with PM & RS & \multicolumn{5}{c|}{DPS} & PCH & DET\\
\cline{3-7}
Module & ~ & M1& M2& M3& M4& M5 & ~ & ~\\
\hline
$B_{0}$~(T) & 4.956 & 5.0& 5.0& 5.0& 5.0& 5.0 & 6.0 & 3.6\\
$\Delta B/B_{0}$~($\%$/30d)~$\leq$&0.011&0.002&0.018&0.001&0.002&0.085 &0.002 &0.002\\
\hline
\end{tabular}
\end{table}
\subsubsection{Magnets with driven mode}
\begin{figure}[t]
\centering 
\includegraphics[width=75.2mm]{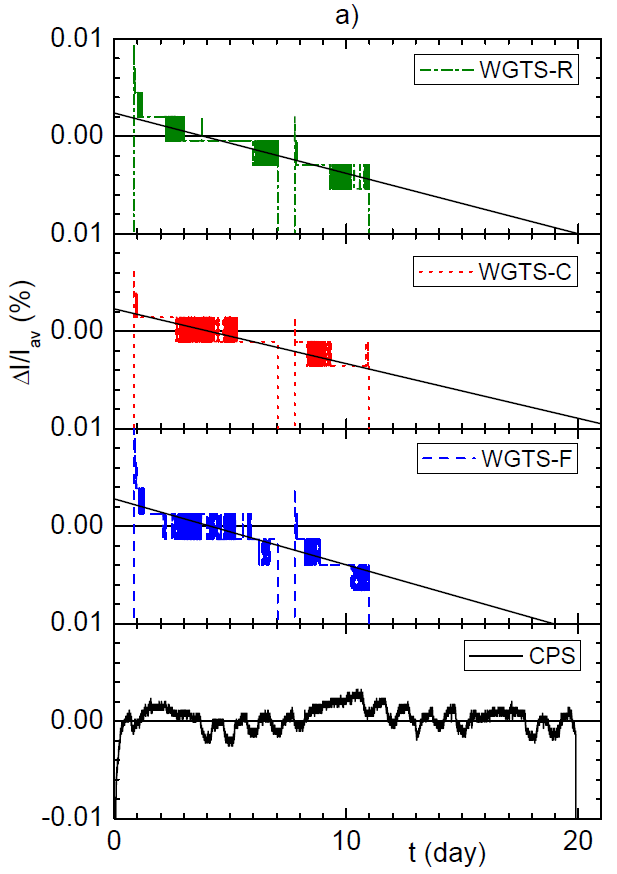}
\includegraphics[width=74.8mm]{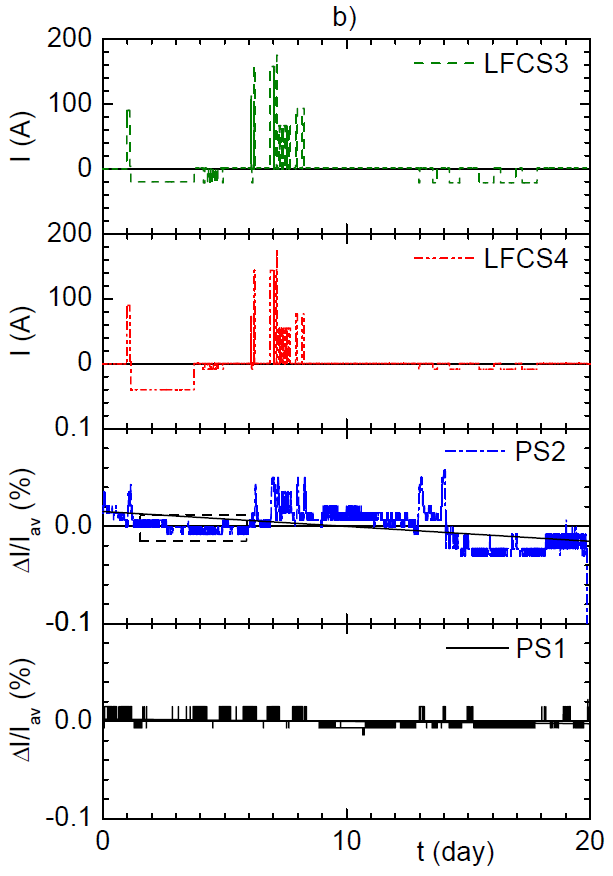}
\caption{Instabilities of the currents at 70$\%$ of the design currents for the magnets operated in driven mode during Run~$\sharp$3. The linear fits on the WGTS data show small drifts over time~(a). At day 7 the currents of the WGTS were lowered to 50$\%$ of the design value. At day 8 it was ramped up to 70$\%$ again. $I_{av}$ for the WGTS currents were taken from day~1 to day~7. $I_{av}$ is shown in table~\ref{tab6}. The instabilities of the PS1 and PS2 currents and the currents of two air coils LFCS3 and LFCS4 are shown in~(b). Data of two representative air coils are shown instead of all 16 air coils, because they are the nearest air coils to the PS2.}
\label{figii1ab}
\end{figure}

The magnetic field stabilities of the magnets with driven mode (DM) are analysed with the data of each DC current transducer in the stabilized power supply units (PSU). Figure~\ref{figii1ab} shows the current instabilities at 70$\%$ of the design currents for the magnets operated in the DM mode during Run~$\sharp$3. The PSUs of the WGTS take a few hours to regulate the set current within 0.01$\%$ after first reaching the set currents. Their current fluctuations are below 0.002$\%$ for long-term operation. However, the WGTS currents show a small drift of < 0.02$\%$ in 30-days, as indicated by the linear fits on the WGTS data (figure~\ref{figii1ab}a), while a very small drift of the CPS current with time is negligible. After operating the WGTS at 50$\%$ for one day and re-charging it back to 70$\%$, the current of the WGTS magnets met the set point within a tolerance of 0.002$\%$ from the initial value (figure~\ref{figii1ab}a). However, the small linear drift of the currents is still noticeable. This small linear drift of the PSUs of the WGTS has to be re-checked later during a longer operation.

Figure~\ref{figii1ab}b shows the current instabilities of the PS1 and PS2 magnets. The current fluctuation of the PS1 is only 0.01$\%$. The same small current fluctuation for the PS2 is shown between day 2 and day 6 in the small dashed box in figure~\ref{figii1ab}b, when the air coils were operated at the currents below 60~A. The fluctuations of the PS2 current up to 0.05$\%$ are compared with operation of the air coils in Run~$\sharp$3. The fluctuations of the PS2 currents occurred, when the currents of the air coils have been changed by more than 60~A until day~8. However, two peaks at days~13~and~14 are not correlated with the air coil currents, because the changes of the air coil currents were below 60~A. The small influence of the air coils on the stability of the PS2 current will be further investigated for long-term operation. Generally one has to wait until the currents have been stabilized after every new current setting or the currents of the air coils have to be changed very slowly. Table~\ref{tab6}~(top) summarizes the results of the current fluctuations and the current drifts of the magnets with the DM mode. The current fluctuations are very small and negligible. The current drifts of the PSU of the WGTS and the magnetic field drifts of all the magnets in table~\ref{tab6} are within the specification (Section~\ref{sec:overview}).
\subsubsection{Stability of magnetic field in the analysing plane of the Main Spectrometer}
The magnetic field stability in the analysing plane of the MS is mostly governed by the stabilities of the  pinch magnet, the PS2 magnet, and the 16 air coils. The instabilities of the magnetic fields of the superconducting magnets are summarized in table~\ref{tab6}. The magnetic field stability of the air coils are defined by the current stabilities of their individual power supplies, because the air coils are always operated in driven mode. The current fluctuations of the power supplies are about 100~ppm. 

However, the magnetic field stability in the analysing plane of the MS can be monitored by the precise flux gate sensors which are installed on the outer surfaces of the MS vessel, as mentioned in section~\ref{sec:ms}. First results of the averaged field drift of 3~($\pm$~1.5)~nT/day was reported for the 0.38-mT setting at the MS~\cite{Erhard2016}.
\subsection{Demonstration of adiabatic electron transport through the whole KATRIN set-up}
\label{sec:beamcheck}
\begin{figure}[t]
\centering 
\includegraphics[width=150mm]{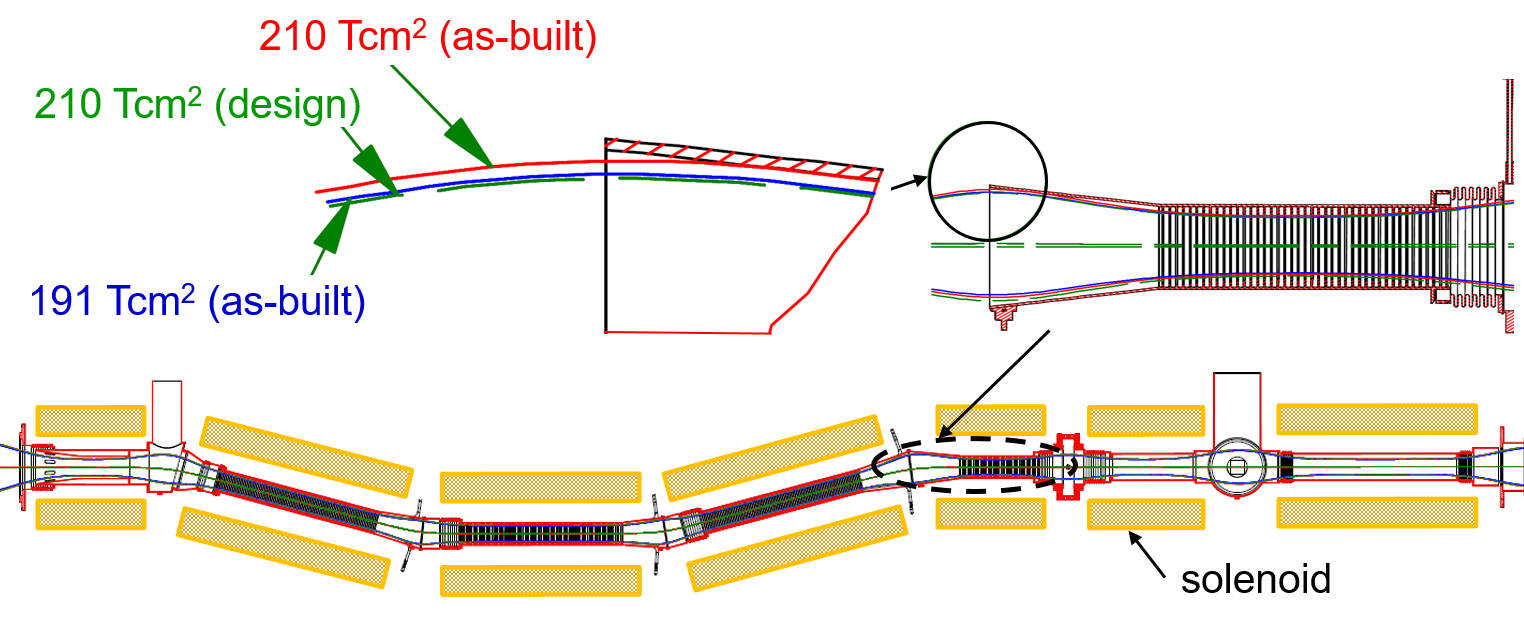}
\caption{Example of the magnetic field calculation for checking beam alignment with the CPS beam tube assembly. The field lines are calculated for the flux of 210~Tcm$^2$ with the design dimensions, for 191~Tcm$^2$, and 210~Tcm$^2$ with the as-built dimensions. The insert shows a vertical cut view (`yz'-plane) relative to the horizontal plane (`zx'-plane in figure~\ref{fig2}b).}
\label{figAlignment}
\end{figure}

The 70-m-long beam tube structures had to be properly aligned relative to the magnetic field lines for the electron transport within the magnetic flux of 191~Tcm$^2$. Therefore, the geometries of the beam tubes were partly machined according to the shapes of the magnetic field lines. They consist not only of straight tube, but also ribs, bellows and cones. Furthermore, a clearance of more than 3~mm of the magnetic field lines relative to the inner structures of the beam tubes has been taken into account for the flux of 191~Tcm$^2$ during the manufacturing and the system assembly. The challenging assembly work of each magnet module and beam tube section had to be accompanied with the measurements of a FaroArm~\cite{FARO} or a laser tracker for each step of the mounting. The measured positions and geometries of the magnets and the beam tube sections have been included in the KATRIN simulation code KASSIOPEIA for magnetic field calculations~\cite{Furse2017, Sack2015, Deffert2017}.  An example of the simulated magnetic field lines is shown for the CPS beam tube assembly in figure~\ref{figAlignment}. The theoretical field calculations with the as-built data shows that the field lines for a flux of 210~Tcm$^2$ hit the surface on a conical section of the beam tube, but the field lines of 191~Tcm$^2$ can still pass the entire beam tube without interference, leaving a reduced clearance of about one millimetre.
\subsubsection{Point-like electron source}
A point-like pencil-beam electron source was deployed to check the beam alignment from the RS to the detector during Run~$\sharp$2 with reduced magnetic fields of about 20$\%$ of the design fields. Thereby, the electrons coming through a 5-mm-diameter small hole in the Rear-Wall of the RS were deflected by the dipole coil pairs (DRx and DRy) at the rear side of the WGTS-R in the x- and y- directions relative to the beam-axis (z-direction). Hence, the point-like electrons could be aligned to each pixel of the detector~\cite{KATRIN2018a}.
\subsubsection{Tritium-like electron source}
\begin{figure}[t]
\centering 
\includegraphics[width=70mm]{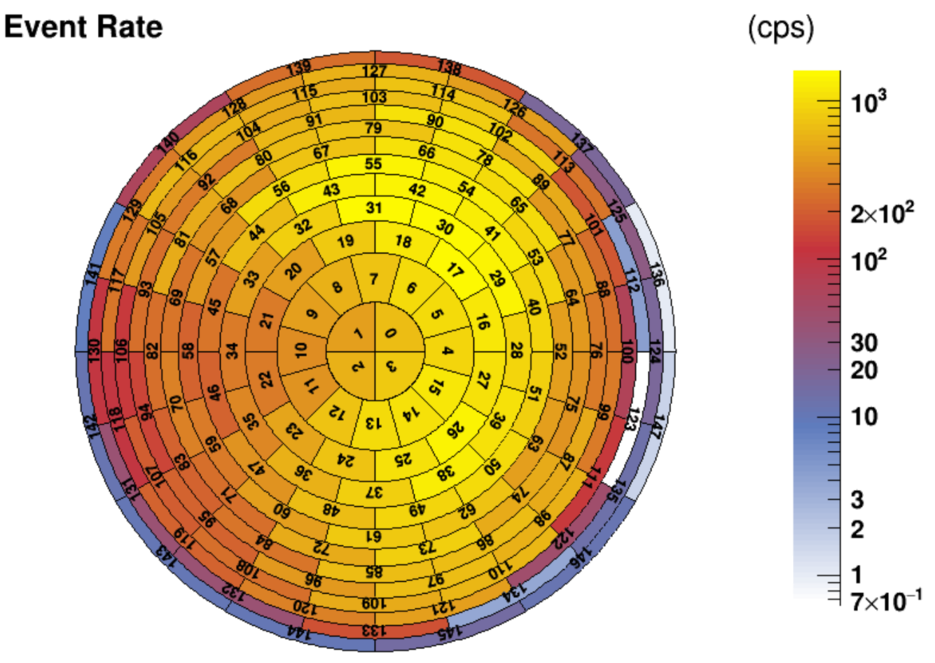}
\includegraphics[width=65mm]{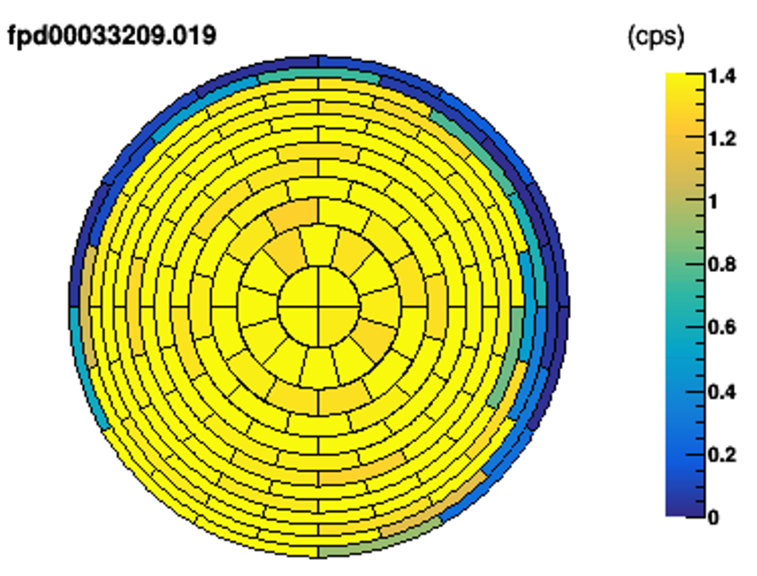}
\caption{Count rates on the detector pixels with 20$\%$ of the design fields during Run $\sharp$2 (left). Count rates on the detector pixels during a gaseous krypton measurement with 70$\%$ of the design fields (Run $\sharp$3) (right).}
\label{figFPD1}
\end{figure}

Unlike the point-like pencil beam source, a ''tritium-like'' electron source generates electrons that fill the flux tube and results in an WGTS emittance that replicates phase-space of the tritium $\beta$-electrons. For the first time on October 14, 2016 KATRIN delivered electrons from the RS to the detector by using a UV-illuminated electron source at the RS (Run $\sharp$1). After the first beam test, the beam alignment was checked in detail using several methods; firstly, by slightly changing the magnetic fields of each magnet group separately, secondly, by adjusting electric potentials on several electric dipoles and monopoles up to several hundred volts. The dipole coil pairs of the WGTS-R were also used to shift the electron beam from the RS relative to the detector's centre (Run $\sharp$2). Figure~\ref{figFPD1}(left) shows an example of the beam alignment check with 20$\%$ of the design fields, confirming that the electrons are guided without interference within the designed magnetic flux from the source to the detector. 
Figure~\ref{figFPD1}(right) shows an example of the electron transport within the magnetic flux of 133.7~Tcm$^2$, corresponding to the 70$\%$ of the design field. The two results are slightly different because of different magnetic field configurations with respect to the source positions. The two preliminary results of the measurements demonstrate the mostly unobstructed transport of the designed magnetic flux of 191~Tcm$^2$ with the complete mechanical assembly~\cite{KATRIN2018a}. Later the beam alignment will be re-checked more precisely with a mono-energetic and angular-selective electron source~\cite{Behrens2017}, gaseous krypton source, and a small amount of tritium in the WGTS-C. 
\section{Lessons learned}
\label{sec:lessons}
\begin{description}
\item[Design of complex magnet systems] It is challenging and time-consuming to manufacture complex magnet systems with beam tubes in one large cryostat in order to achieve many technical requirements for physics. It is recommended to separate the magnet systems from other systems in design, if possible. The former DPS2-F magnet system was designed in one cryostat because of the 77-K beam tube temperature requirement. However, its beam tube design was revised to allow room temperature operation by adding additional TMPs for more tritium pumping capacity. Therefore, the new DPS magnets could be manufactured with five single magnets in much shorter delivery time.
\item[Accessibility of critical key components] Critical key components of the two large magnet systems (the WGTS and the CPS) had to be re-considered with respect to their reliability and accessibility for a potential repair. Conventional design for small single magnets cannot be properly adapted for a large complex magnet system without special design for the key components. Critical components, such as the persistent switch heater and cold by-pass diodes, should not be located in an unreachable areas to allow for practical access in the event if repairs are required. Otherwise, their repair later may be practically not possible. Therefore, we removed the persistent current switch by changing the operation mode to the driven mode at the WGTS and the CPS in design phase. In addition, we improved the design of the two large magnet systems with separate vessels for the cold by-pass diodes, where they are easily accessible from outside at warm condition.
\item[Optimisation of thermal cycling of single magnets] Leak tightness of the magnet cryostat and its insulation vacuum condition are important prerequisites among others especially for small single magnets which have O-rings or indium-seals, because their leak tightness can be reduced by rapid thermal cycling. Moisture can also penetrate into  system. Therefore, after every thermal cycle the insulation vacuum condition has to be re-checked. The time of the cool-down and warm-up of the system has to be optimised to avoid any thermal stress on the seals.
\item[Optimisation of cool-down procedure of recondensing magnets] A typical cool-down procedure of a such small helium bath cryostat is a fast pre-cooling down of the coil by liquid nitrogen, afterwards blowing out by dry-nitrogen, and filling in liquid helium from around 120~K within a week. However, this kind of shock-like fast cooling can cause thermal stress on the weak connection elements like sealing material and flanges, resulting in an unwanted small leak. This can consume more efforts and significant time for re-cooling, especially if the magnets are confined within the secondary containment of the tritium enclosure. Therefore, the cooling procedure of the small recondensing magnets has been optimised for a reliable system cooling. Direct conduction-cooling by the PT415 cryocooler from room temperature has been tested with two magnets of the DPS and optimised by automatic control of the overpressure in the helium chamber with ``red-y smart pressure controller GSP'' from V{\"o}gtlin~ Instruments~AG.\footnote{Product information of red-y smart smart pressure controller: https://www.voegtlin.com/data/329-2060$\_$en$\_$infosmartpressure.pdf} The controller was able to keep a small overpressure of 8~kPa in the helium chamber during the cool-down by automatically regulating the gas flow into the helium chamber. The cool-down time down to about 50~K took about 10~days. Afterwards the liquid helium from a Dewar could be transferred into the magnet chamber. During the filling, loss of liquid helium could be significantly reduced. The test demonstrated a possibility for minimizing the thermal stress in the system and a potential risk of a leak.
\item[PM mode of single magnets]
As soon as the single magnets were swithed to the PM mode, their power supplies have to be completely ramped down to zero. It avoids a small conduction heat input to the magnet circuit. Otherwise, persistent currents can drift faster, depending on the thermal condition.
\end{description}
\section{Conclusion}
\label{sec:conclusion}
All KATRIN superconducting magnets were successfully operated for about two weeks continuously during Run~$\sharp$2 for the beam alignment and other measurements at 20$\%$ of the design fields and during the first krypton measurements at 70$\%$ of the maximum design fields (Run~$\sharp$3). 

The beam alignment was successfully confirmed for an mostly unobstructed electron transport for the designed magnetic flux of 191~Tcm$^2$. It will be re-checked with the complete system by an e-gun at the RS and finally with the gaseous krypton and gaseous tritium in the source. 

Both the field drift of the single magnets in persistent current mode and the current fluctuations of the two large magnets in driven mode were shown to be well within the limits of the stringent KATRIN requirements. It will be re-checked during each 60-days-run of the standard operation.

The nominal magnetic flux density of the KATRIN experiment is currently reduced to 70$\%$ of the design fields in order to operate the experiment in a safe region without the risk of a quench. The nominal magnetic flux density can be re-defined at a later time, considering all other experimental conditions.

The magnet safety systems (MSS) were successfully commissioned with the WGTS and the CPS. It can detect not only a quench in an individual magnet system but also a quench in their adjacent magnets. In case of a quench validation, the MSS triggers the external dumping unit to discharge the magnets according to pre-defined rules. It helps one not only to effectively protect the magnets but also to minimize the loss of a huge amount of liquid helium and thus the cryogenic recovery time.

The KATRIN experiment will be fully operational after the remaining tritium circuits are completely installed in 2018. All components have to be fully commissioned before the official approval of tritium operation. The first tritium run is scheduled in June, 2018.\\


\acknowledgments
We acknowledge the support of Helmholtz Association (HGF), Ministry for Education and Research BMBF (05A17PM3, 05A17PX3, 05A17VK2, and 05A17WO3), Helmholtz Alliance for Astroparticle Physics (HAP), and Helmholtz Young Investigator Group (VH-NG-1055) in Germany; Ministry of Education, Youth and Sport (CANAM-LM2011019), cooperation with the JINR Dubna (3+3 grants) 2017-2019 in the Czech Republic; and the Department of Energy through grants DE-FG02-97ER41020, DE-FG02-94ER40818, DE-SC0004036, DE-FG02-97ER41033, DE-FG02-97ER41041, DE-AC02-05CH11231, and DE-SC0011091 in the United States.

The authors would like to thank D.~Hagedorn from CERN and A.~Herv\'{e} from the University of Wisconsin for their valuable consultations and R.~Gehring and M.~Noe with ITEP of KIT and A. Kosmider with IKP of KIT for their early contributions.







\end{document}